\documentclass[reprint,preprintnumbers,showpacs,showkeys,
prl
,amssymb,amsmath,amsfonts,]{revtex4-2}

\usepackage[utf8]{inputenc}
\usepackage[T1]{fontenc}
\usepackage{graphicx}
\usepackage{mathrsfs}
\usepackage{bm}
\usepackage[dvipsnames]{xcolor}
\usepackage{mathtools}
\usepackage{hyperref}
\usepackage{tcolorbox}
\usepackage{float}
\usepackage{textcomp}
\usepackage{flafter}

\makeatletter
\def\bibstyle#1{}
\def\bibdata#1{}
\makeatother

\date{\today}

\begin{document} 

\title{Repulsive Inverse-Distance Interatomic Interaction from \\Many-Body Quantum Electrodynamics}

\author{Loris Di Cairano}              
\affiliation{Department of Physics and Materials Science, University of Luxembourg, L-1511 Luxembourg City, Luxembourg}

\author{Matteo Gori}              
\affiliation{Department of Physics and Materials Science, University of Luxembourg, L-1511 Luxembourg City, Luxembourg}

\author{Reza Karimpour}              
\affiliation{Department of Physics and Materials Science, University of Luxembourg, L-1511 Luxembourg City, Luxembourg}

\author{Alexandre Tkatchenko}            
\email{alexandre.tkatchenko@uni.lu}
\affiliation{Department of Physics and Materials Science, University of Luxembourg, L-1511 Luxembourg City, Luxembourg}

\begin{abstract}
Interactions between objects can be classified as fundamental or emergent. Fundamental interactions are either extremely short-range or decay inversely with the separation distance, such as the Coulomb potential between charges or the gravitational attraction between masses. In contrast, emergent quantum van der Waals (vdW) and Casimir interactions decay considerably faster ($R^{-6}$ or $R^{-7}$) with distance $R$. Here we apply perturbative quantum electrodynamics (QED) to a many-body (MB) system of atoms modeled as charged harmonic oscillators, and reveal a persistent inverse-distance MB-QED interaction stemming from the coupling between virtual photons and molecular plasmons in the non-retarded regime. This interaction, scaling with the third power of the fine-structure constant, is reminiscent of the Lamb shift for a single atom. Although weaker than vdW forces, this MB-QED $R^{-1}$ interaction may substantially surpass gravitational attraction in future experiments probing quantum gravity at microscopic scales.
\end{abstract}


\maketitle

The quantum vacuum, as the ground state of interacting fields, gives rise to both fundamental forces between charges and emergent forces between neutral systems~\cite{casimir1948attraction,milonni2013quantum,craig1998molecular,tkatchenko2015current}. 
Quantum electrodynamics (QED), regarded as one of the most successful and precise theories in physics, produced an exquisitely accurate description of the interactions between charged particles and the electromagnetic field (EMF)~\cite{salam2009molecular,salam2016non,craig1998molecular,buhmann2013dispersion,milonni2013quantum,passante2018dispersion}. However, only recently has QED been extended to treat realistic molecules and materials composed of many particles \cite{gori2023second,karimpour2022quantum,karimpour2022molecular}. This is a very productive and intriguing direction of research that has yielded seminal results, especially for matter in cavities and under a strong coupling regime~\cite{flick2017atoms,riso2022molecular,forn2019ultrastrong,ashida2021cavity,xiang2024molecular,haugland2020coupled,hertzog2019strong,mandal2023theoretical}. 

Much less is known about the weak-coupling effects of QED for many-body quantum matter; in particular, can one expect emergent interactions when coupling many interacting atoms or molecules with the EMF? Conventional molecular QED that treats atoms as isolated dipoles perturbatively coupled to the EMF has been very successful. However, a rigorous and unified treatment of electromagnetic interactions between multiple molecules has remained intractable~\cite{salam2009molecular}. Additionally, at short distances and minimal coupling, the scalar charge-charge interaction is stronger than the dynamic charge-EMF interaction; thus, perturbative QED must be used with care. 
In this work, we combine the many-body dispersion (MBD) formalism~\cite{tkatchenko2012accurate,distasio2014many}---which captures the collective electronic response in molecular systems to infinite order---with perturbative QED, 
resulting in a unified theory we refer to as many-body QED (MB-QED). 
In the MBD framework, atoms are modeled as quantum Drude oscillators (QDOs) coupled via dipolar interactions. MBD is fully general because it provides a transferable and accurate description of collective electronic fluctuations and is applicable in chemistry, biology, nanoscale systems, and condensed matter~\cite{woods2016materials,hermann2017first}.

%
\begin{figure*}[!htbp]
  \centering
\includegraphics[height=18cm, width=18cm,keepaspectratio]{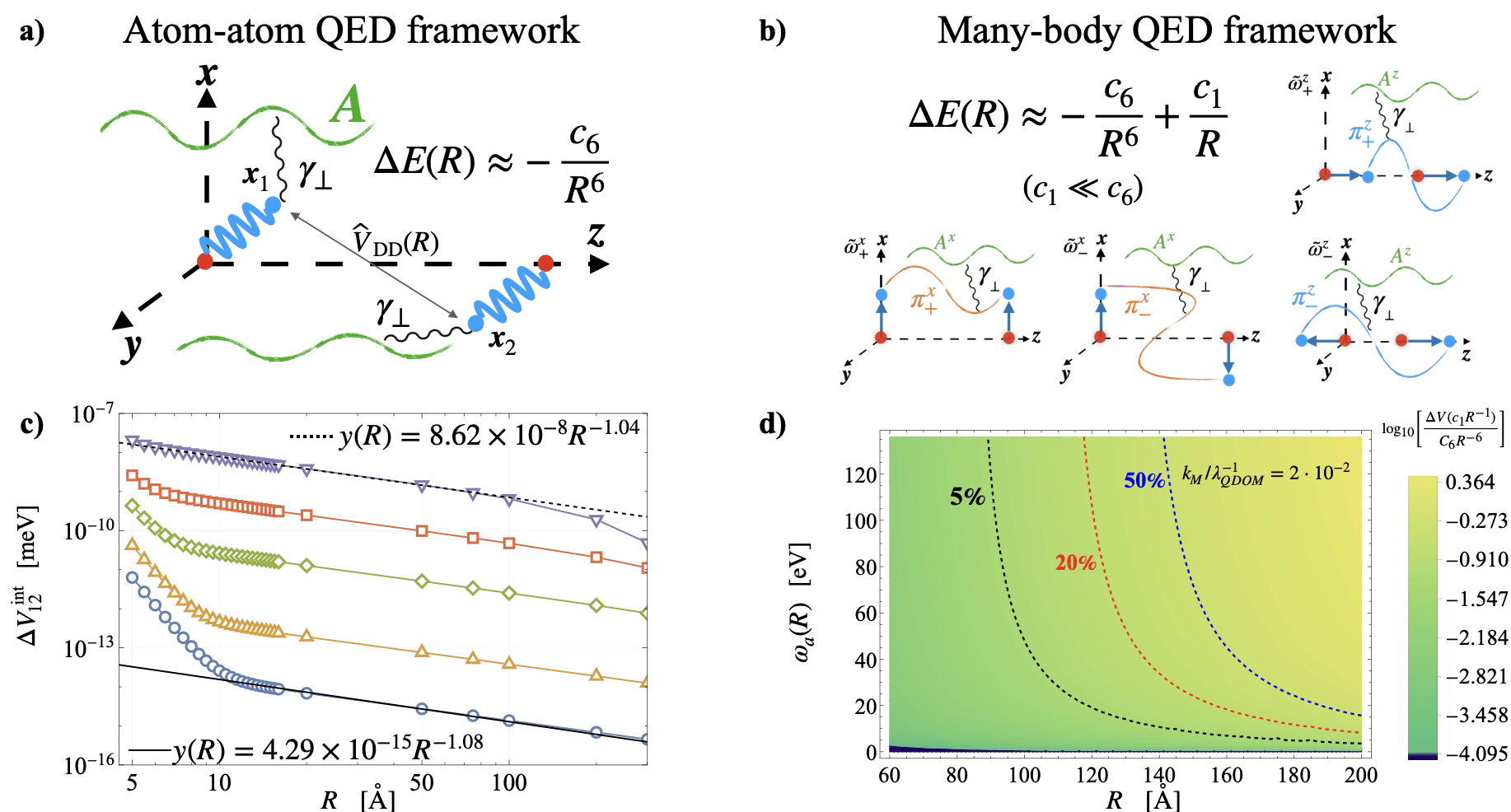}
\caption{\textbf{a)} Two quantum Drude oscillators (QDOs) interacting with one another via dipole-dipole coupling $\widehat{V}_{\mathrm{DD}}$, and with the electromagnetic field (EMF) through the minimal coupling term $\widehat{\bm{p}}_{a} \cdot \widehat{\bm{A}}(\bm{R}_a)$ that results in the exchange of transverse photons $\gamma_{\perp}$.
\textbf{b)} QDOs normal modes resulting from diagonalization of the many-body dispersion (MBD) Hamiltonian, \textit{i.e.} $\widehat{H}_{\mathrm{MBD}}=\widehat{H}_{\rm QDO}^{(1)}+\widehat{H}_{\rm QDO}^{(2)}+\widehat{V}_{\mathrm{DD}}$ (see SM for the diagonalization). There are six normal modes, including two longitudinal modes ($\pi_\pm^z$) and four transverse modes ($\pi_\pm^x$, and two other modes with $\pi_\pm^y$ that are not shown here). 
These modes interact with the EMF, exchanging transverse photons $\gamma_{\perp}$, via $\widehat{H}_{\text{int}}=-\sum_{L,i}\widehat{\pi}_{L,i}  \mathcal{A}_{L,i}$ as in Eq.~\eqref{def:interaction_term}. 
\textbf{c)} Interaction energy $\Delta V^{\text{int}}_{12}$ defined in Eq.~\eqref{eq:DeltaV_int} as a function of the inter-QDO distance $R$ for different values of the cutoff ratio $k_M/\lambda_{\rm QDOM}^{-1}$, where $\lambda_{\rm QDOM} = \max_a \sqrt{\hbar/2m\omega_a}$ (see main text). The violet triangles, red squares, green diamonds, orange triangles, and blue circles correspond respectively to $k_M/\lambda^{-1}_{\rm QDOM}=2 \cdot 10^{-2},\ 10^{-2},\ 5 \cdot 10^{-3},\ 2 \cdot 10^{-3}$, and $10^{-3}$. The dashed black line shows the fit $y(R) = 8.62 \times 10^{-8} R^{-1.04}$, while the solid black line corresponds to $y(R) = 4.29 \times 10^{-15} R^{-1.08}$. Notice that the deviations from the $1/R$ behavior at short range and low frequencies arise from the higher-order terms, particularly the $R^{-9}$ term in \eqref{def:R_behavior}, which dominates over $1/R$ for smaller values of the cut-off wavenumber $k_M$ (see Sec. V in the SM for a detailed analysis).
%
\textbf{d)} Contour plot of $\log_{10}[\Delta V(c_1 R^{-1}) / c_6 R^{-6}]$ as a function of QDO frequencies and interatomic distance $R$. This illustrates the crossover between the many-body QED correction and the van der Waals interaction across spatial and frequency scales.} 
  \label{fig:all}
\end{figure*}

The well-established atom--atom QED approach assumes that atoms are polarizable entities that interact perturbatively with each other and with the EMF (see panel \textbf{a} of Fig.~\ref{fig:all})~\cite{salam2009molecular,salam2016non,craig1998molecular}.
This approach ignores interatomic correlations and high-frequency field modes. This logic underpins the vast majority of EMF force calculations, except for rare case examples where non-perturbative approaches were considered ~\cite{crisp1991interaction,renne1971retarded, compagno1995atom,ciccarello2005exactly}.

In realistic molecules and solids, interacting atoms form a correlated electronic system. Strong interatomic bonding and many-body correlations must be taken into account before coupling to the EMF vector potential. We do this and reveal a repulsive interaction in the near-zone that scales as $R^{-1}$ between neutral atoms arising from the coupling between molecular plasmons and virtual photons. 



We developed a general MB-QED framework for a system of QDOs coupled with the EMF (see Supplemental Material (SM) for the derivation). However, our main findings can already be explained by the example of two coupled atoms forming molecular plasmons, which we present in what follows. We consider the Hamiltonian describing the QDOs system in interaction with the EMF:
\begin{equation}
\label{def:H-QDO-EMF}
    \widehat{H}_{\text{QDO-QED}}= \sum_{a=1}^2 \widehat{H}_a + \widehat{V}_{\text{DD}}(R) 
    + \widehat{H}_{\text{EMF}}
    +\widehat{H}_{\text{int}}\ .
\end{equation}
The term $\widehat{H}_{a}=|\widehat{\bm{p}}_a|^2/2+m_a\omega^2_{a} |\widehat{\bm{r}}_{a}|^2/2$ describes the QDO system parametrized by effective mass $m_a$, charge $q_a$ and frequency $\omega_a$, and is represented by the canonical operators $\widehat{\bm r}_a=(\widehat{x}_a,\widehat{y}_a,\widehat{z}_a),\,\widehat{\bm p}_a=(\widehat{p}_{a,1},\widehat{p}_{a,2},\widehat{p}_{a,3})$ with $a=1,2$, located on the $z$-axis at $\bm{R}_1=R_1\bm{e}_z$ and $\bm{R}_2=R_2\bm{e}_z$ with mutual distance $ R= |R_2- R_1|$. The oscillators are coupled through the dipole-dipole potential $\widehat{V}_{\rm DD}=q_1q_2e^2(\widehat{\bm{r}}_1\cdot \widehat{\bm{r}}_2-\widehat{z}_1 \widehat{z}_2)/4\pi\epsilon_0 R^3$. The quantum energy levels for each QDO are $E^i_{a}=\hbar\omega^i_{a}(m^i+1/2)$ where $m^i$ is the quantum number in the $i$-th spatial direction. 

The EMF, represented by the vector potential, $\widehat{\bm{A}}$, is considered in Coulomb gauge $\nabla\cdot\bm{A}=0$ and in dipole approximation, i.e., assuming 
$\widehat{\bm{A}}(\bm{R}_{a}+\bm{r}_a)\simeq \widehat{\bm{A}}(\bm{R}_{a})$. In so doing, $\widehat{H}_{\rm EMF}$ is the free Hamiltonian of the EMF, which is the sum of the squares of the electric and magnetic fields, while the QDO-EMF interaction term takes the form $\widehat{H}_{\rm int}=\sum_a (-q_a\widehat{\bm{p}}_a\!\cdot\!\widehat{\bm{A}}(\bm R_a)/m_a+q_a^2\widehat{\bm{A}}^2(\bm{R}_a)/2m_a)$, where
$ \hat{A}_j(\bm x)\!=\!\sum_{\bm k,\lambda}\sqrt{\hbar(2\varepsilon_0 c k V)^{-1}} \epsilon_{j,\lambda}(\bm k)(\hat a_{\bm k\lambda}\,e^{i\bm k\cdot\bm x}\!\!+c.c.)$, $\bm{k}$ is the wave vector, $\{\bm{\epsilon}_{\lambda}(\bm{k})\}$ are the polarization vector basis, and $\widehat{a}_{\bm{k}\lambda}$ and $\widehat{a}^{\dagger}_{\bm{k}\lambda}$ are, respectively, the photon annihilation and creation operators.

The diamagnetic $\bm{A}^2$ term does not play a relevant role in MB-QED (see Sec. III.A in the SM). Its first-order contribution is purely local and independent of the interatomic distance, while its second-order effect is of higher order in the charge compared to the $q^2$ correction arising from the paramagnetic coupling. Therefore, it does not contribute to the interaction energy and can be safely neglected within the MB-QED framework.

We treat the dipole–dipole interaction exactly by diagonalizing the coupled Hamiltonian \(\widehat{H}_0 = \widehat{H}_1 + \widehat{H}_2 + \widehat{V}_{\rm DD}\). 
This diagonalization defines collective normal modes---molecular plasmons---whose renormalized frequencies 
$\widetilde{\omega}_{L,i}$ already include the full instantaneous electronic correlation between the atoms.
The transformation from individual to collective operators is realized through the orthonormal change of basis $\widehat{\bm{X}}_{L}=\sum_{a=1}^2\bm{C}_{aL}\,\widehat{\bm{r}}_{a}$, $\widehat{\bm{\pi}}_{L}=\sum_{a=1}^2\bm{C}_{aL}\,\widehat{\bm{p}}_{a}$,
with 
$\bm{C}_{1-}=\bm{C}_{1+}=\bm{C}_{2+}=\tfrac{1}{2}\,1\!\!1$ and
$\bm{C}_{2-}=-\tfrac{1}{2}\,1\!\!1$. In these collective coordinates, the QDO-QED Hamiltonian in Eq.~\eqref{def:H-QDO-EMF} writes
\begin{equation}\label{def:MB-QED-hamiltonian}
    \widehat{H}_{\rm MB-QED}=\widehat{H}_{\rm MBD}+\widehat{H}_{\rm int}+\widehat{H}_{\rm EMF}
\end{equation}
where the interaction reads
\begin{equation}\label{def:interaction_term}
    \widehat{H}_{\rm int}=-\sum_{L,i} \widehat{\pi}_{L,i}\;\widehat{\mathcal{A}}_{L,i},\;\; \widehat{\mathcal{A}}_{L,i}=\sum_{a,j} \frac{q_a(C_{aL})_{ij}}{\sqrt{m_a}}\widehat{A}_i(\bm{R}_a)\,.
\end{equation}
The molecular subsystem assumes a diagonal form
\begin{equation}
    \widehat{H}_{\rm MBD} = \frac{1}{2}\sum_{L=\pm}\sum_{i=1}^{3}\Big(\widehat{\pi}_{L,i}^2+\widetilde{\omega}^2_{L,i}\widehat{X}_{L,i}\Big)\;.
\end{equation}
representing the system of six normal-mode QDOs, each characterized by its corresponding normal-mode frequency $\widetilde{\omega}_{(L,i)}$~(see panel \textbf{b} of Fig.~\ref{fig:all}). The ground-state energy of this collective system contains not only the ground-state energy of isolated QDOs (described by $\sum_{a=1}^2\widehat{H}_a$), but also the distance-dependent interaction energy that arises from the instantaneous and anisotropic dipole–dipole coupling between them. This interaction energy contains an infinite series of London dispersion and polarization contributions, starting with the leading order $R^{-6}$ term~\cite{tkatchenko2013interatomic}. This implies that the polarizability of the QDO dimer is non-additive at any separation where the dipole--dipole potential works. 

Within this collective formulation, the leading radiative energy arises from the second-order interaction between each normal mode and the quantized EMF
\begin{equation}\label{def:second_order_perturbation}
    \Delta E^{\rm MB-QED} = - \sum_{i,L}\sum_{\bm{k}\mu} \dfrac{|\langle \tilde{1}_{i,L};1_{\bm{k},\mu}|\widehat{H}_{\rm int}| 0_{i,L};0_{\bm{k}\mu}\rangle|^2}{\hbar c k +\hbar \widetilde{\omega}_{i,L}}
\end{equation}
A cumbersome calculation (see Sec. IV in the SM) recasts the radiative correction above into an intermediate form
\begin{align}\label{def:intermediate-DeltaE}
    \Delta E^{\rm MB\text{-}QED}=&-\frac{\hbar}{4\varepsilon_0 c}
    \sum_{L,j}\!\int\!\frac{dk\,d\Omega_{\bm{k}}}{(2\pi)^3}
    \frac{k\widetilde\omega_{Lj}}{(c k+\widetilde\omega_{Lj})}\\
    \times
    \sum_{a,b}\sum_{i,\ell}
    &\frac{q_a q_b}{\sqrt{m_a m_b}}
    (C_{aL})_{ij}(C_{bL})_{\ell j}
    P_{i\ell}(\bm k)\,e^{i\bm k\cdot(\bm R_a-\bm R_b)} \nonumber.
\end{align}
Note that we have defined $P_{ij}(\bm{k}):=\sum_{\lambda} \epsilon_{i,\lambda}(\bm k)\epsilon_{j,\lambda}(\bm k)=\delta_{ij}-k_ik_j/k^2$. The integral naturally separates into a self-energy part $\Delta U_a^{\mathrm{self}}(k_M)$ for $a=b$ and an interaction part $\Delta V_{12}^{\mathrm{int}}(R,k_M)$ for $(a\neq b)$, corresponding respectively to the dressing of each collective mode and the photon-mediated coupling between them.

The self-energy term carries the usual ultraviolet divergence of QED. We choose the renormalization scheme such that the reference configuration is represented by a set of dynamically isolated collective modes, rather than isolated atoms. This choice is the most appropriate to enforce consistency with the MBD Hamiltonian $\widehat{H}_{\rm MBD}$ since it naturally preserves its normal-mode structure (see Sec. IV in SM). 
The reference shift for each mode reads
\begin{equation}
    \Delta E_a^{\rm free}(k_M)=-\frac{\hbar}{12\pi^2\varepsilon_0 c}
    \frac{q_a^2}{m_a}
    \sum_{L,j}\mathcal{C}_{a,L}\!\!\int_0^{k_M}\!\!\!dk\,\frac{k\,\widetilde\omega_{Lj}}{c k},
\end{equation} 
where $\mathcal{C}_{a,L}:=\mathrm{Tr}(C_{aL}^{T}C_{aL})$.

Subtracting this term from Eq.~\eqref{def:intermediate-DeltaE} removes the non-physical divergence and yields a finite, physically meaningful correction to the self-energy, i.e., $\Delta U_a^{\mathrm{self}}(k_M)-\Delta U_a^{\mathrm{free}}(k_M)$, that gives 
\begin{align}
\label{eq:DeltaU_self}
        \Delta U_a^{\rm self,\,ren}(k_M)=&\frac{\hbar}{12\pi^2\varepsilon_0 c^3}
    \frac{q_a^2}{m_a}\nonumber\\
    &\times\sum_{L,j}\mathcal{C}_{a,L}\widetilde\omega_{Lj}^2
    \ln\!\!\left[\frac{c k_M+\widetilde\omega_{Lj}}{\widetilde\omega_{Lj}}\right].
\end{align}
This self-energy does not have any distance-dependent contributions to the energy. In contrast, the remaining finite contribution defines the effective interaction between collective excitations:
\begin{equation}
\label{eq:DeltaV_int}
\begin{split}
    \Delta V_{12}^{\rm int}(R,k_M)&=-\frac{\hbar}{4\pi^2\varepsilon_0 c}
    \frac{2\,q_1 q_2}{\sqrt{m_1 m_2}}\\
    \times\!\sum_{j}\int_0^{k_M}\!\!\!\!&\bar f(kR)\left[\frac{k\,\widetilde\omega_{+j}}{c k+\widetilde\omega_{+j}}-\frac{k\,\widetilde\omega_{-j}}{c k+\widetilde\omega_{-j}}
\right]dk
\end{split}
\end{equation}
where $\bar{f}=(f_{\parallel}+2f_{\perp})/2$ and $f_\perp(x)=(x\cos x+(x^2-1)\sin x)/x^3$ and $f_\parallel(x)=2[\sin x-x\cos x]/x^3$. This term encapsulates the full radiative coupling between molecular plasmons. All terms are finite, parameter-free, and directly determined by the microscopic parameters of the QDO model.


The inverse-distance behavior of the interaction energy $\Delta V_{12}^{\mathrm{int}}(R,k_M)$ emerges when the structure of Eq.~\eqref{eq:DeltaV_int} is examined in the near-zone regime. In this limit, both the MBD eigenstates and the eigenfrequencies can be perturbatively expanded in powers of the dipole-dipole coupling constant $\gamma=\mathcal{A}_{0}/R^{3}$, around the corresponding values for the uncoupled QDOs. Moreover, the geometric functions $f_{\parallel}(kR)$ and $f_{\perp}(kR)$ can be expanded in series around $\tau = kR=0$. After expanding the integrand in Eq.~\eqref{eq:DeltaV_int} to third order in $\gamma$ and $\tau$, performing the integration over $k$, and restoring the explicit dependence on $R$ (see Sec. V in SM for further details), one obtains a compact expression that captures the spatial dependence of the QED contribution to the interatomic interaction potential
\begin{equation}\label{def:R_behavior}
\Delta V_{12,\mathrm{app}}^{\mathrm{int}}(R,k_M) = \Delta V \left[\frac{c_1(k_M)}{R} + \frac{c_7(k_M)}{R^7} + \frac{c_9(k_M)}{R^9} \right]
\end{equation}
where $\Delta V$ denotes a reference energy scale and where the coefficients are reported in the SM. Adopting atomic units ($[c]_{a.u.} = \alpha_{\rm fsc}^{-1}$), we find $\Delta V  = \alpha^3_{\rm fsc} E_{h} [E_M]_{a.u.} [(\mathcal{A}_{0,1} \mathcal{A}_{0,2})^{1/2}]_{a.u.} [(E_1 E_2)^{3/2}]_{a.u.}/\pi$ with $E_{M} = \hbar c k_{M}$, $E_h$ the Hartree energy scale, and $\alpha_{\rm fsc}$ is the fine-structure constant.

Remarkably, when compared to the explicit numerical solution of Eq.~\eqref{eq:DeltaV_int}, the analytic prediction remains valid over a wide distance range ($\sim$5--100~\r{A}) even for relatively large values of $k_M$. The MB-QED interaction scales with $\alpha^3_{\rm fsc}$, which is reminiscent of the Lamb shift for a single atom.
We observe that the only QED contribution that decays slower than the standard $R^{-6}$ vdW interaction is the term $c_{1}\,\Delta V / R$, with $c_1(k_M)>0$, leading to a repulsive long-range interaction. This term originates by taking the first order in $\gamma\propto R^{-3}$ and the second order in $\tau\propto R^{2}$ in the series expansion used to derive Eq.\eqref{eq:DeltaV_int}.
This provides an interpretation of the $R^{-1}$ term as arising from elementary 
quanta-exchange processes between QDOs and the EMF. 

Within the second-quantized
 MBD formalism of \cite{gori2023second}—which employs a Bogoliubov 
transformation between the atomic QDO and the MBD creation/annihilation operators—the 
eigenstates of the MBD Hamiltonian can be expressed as linear superpositions of free 
atomic QDO eigenstates, expanded in powers of $\gamma$. In this framework, dipole–dipole MBD interactions induce virtual currents in the atomic QDOs, which subsequently 
couple to the EMF, effectively acting as excited dipolar 
currents. Our result can thus be interpreted in analogy with previous works on the 
emergence of long-range interactions between excited dipole states in both classical 
\cite{preto2015possible} and quantum electrodynamics 
\cite{gomberoff1966long,barcellona2016van}. 


Our approximate analytical derivation is fully confirmed by explicit numerical 
evaluation of the radiative integral in Eq.~\eqref{eq:DeltaV_int}. We take the argon 
dimer as a prototypical example and calculate $\Delta V_{12}^{\text{int}}$ by 
numerical integration. We adopt a regularization scheme in which the cutoff $k_{\rm 
M}$ is defined by the inverse of the largest characteristic length scale of the atomic 
QDOs. Specifically, we set \(k_{\rm M} = \eta \lambda_{\rm QDOM}^{-1}\), with $0 < 
\eta \leq 1$, and $\lambda_{\rm QDOM} = \max_{a=1,2} \sqrt{\hbar/2m_a\omega_a}$. Note that $\eta 
\leq 1$ comes from a QDO describing the atomic response; thereby, we exclude EMF 
modes with wavelengths short enough to resolve the internal structure of a QDO. The 
results of this numerical computation are reported in Fig.~\ref{fig:all}\textbf{c} and 
compared to the analytic expansion. We find excellent agreement between both methods. 
This repulsive component becomes particularly relevant in the intermediate regime 
between the non-retarded and retarded limits, where it can reach up to $\sim$10\% of 
the total dispersion energy (Fig.~\ref{fig:all}\textbf{d}) for an argon dimer. The excitation frequency of molecular plasmons controls the overall strength of the MB-QED interaction versus the standard vdW interaction. For example, tightly bound core electrons would contribute more strongly to the MB-QED effect. In contrast, lower excitation frequencies increase the spatial range of the near zone, making the MB-QED interaction longer and more pronounced.  


A complete understanding of our result would require working out a non-perturbative approach to field-matter coupling, for example, extending the method proposed by Renne~\cite{renne1971retarded}. 
Indeed, many-body quantum mechanics shows that one needs at least fourth-order wavefunctions to reflect the effect of second-order energies on quantum states~\cite{szabo1996modern}.
At this stage, we strive to propose possible interpretations of the derived inverse-distance interaction by making an analogy with other $1/R$ terms between point-like particles. The MB-QED effect can be interpreted as arising from a scalar isotropic potential generated by an effective source. Since the only fundamental long-range $1/R$ interactions are electrostatic and gravitational, we assign identical effective charges and masses such that the associated $1/R$ potentials reproduce the QED energy correction.
In atomic units, where the Coulomb potential reads $V(R)=q_1 q_2 / R$, the MB-QED interaction takes the form
\(
\Delta V_{12}^{\text{int}}(R)=\alpha_{\mathrm{fsc}}^{3}
\sqrt{\mathcal{A}_{0,1}\mathcal{A}_{0,2}}\,(E_1E_2)^{3/2}/\pi R,
\)
where $\mathcal{A}_{0,a}$ is the static polarizability (in $a_0^3$) and $E_a=\hbar\omega_a$ the QDO excitation energy (in Hartree). 
By identification with the Coulomb term, we define a per-object \emph{effective charge}
\( q_{\mathrm{eff},a}=\alpha_{\mathrm{fsc}}^{3/2}\,
\mathcal{A}_{0,a}^{1/4}\,E_a^{3/4}/\sqrt{\pi},
\)
such that $\Delta V_{12}^{\text{int}}(R)=q_{\mathrm{eff},1}\,q_{\mathrm{eff},2}/R.$
Unlike the elementary electron charge, $q_{\mathrm{eff}}$ is an 
emergent property of correlated matter; it condenses measurable 
quantities---polarizability and excitation energies---into an 
effective coupling to the quantum vacuum.
The same $1/R$ law can also be written in a gravitational form, as if the vacuum modes carried a \emph{effective imaginary mass} renormalized through their coupling to matter. 
As shown in Refs.~\cite{szabo2022four,goger2024four}, the static polarizability of QDOs is proportional to the mass $\alpha^{\text{GKT}}_a=C m_a q_a^2$ with $a=1,2$ and where $C=4L^4/9\hbar^2$ is a constant depending on the characteristic length $L$ of the QDO. Now, in Eq.~\eqref{def:R_behavior}, $\Delta V^{\text{int}}_{12,\text{app}}$, the terms $\Delta V$ is proportional to $\mathcal{A}_{0,1} \mathcal{A}_{0,2})^{1/2}$ and similarly $c_1(k_M)=(\mathcal{A}_{0,1} \mathcal{A}_{0,2})^{1/2}f(k_M)$ where $f(k_M)$ is a renormalization factor appearing in Eq.~(65) in the SM. Therefore, we get $\Delta V^{\text{lead}}(R)=\Delta V c_1/R\propto \mathcal{A}_{0,1} \mathcal{A}_{0,2}/R$, and using the relation $\alpha^{\text{GKT}}$, we obtain $\propto m_1m_2/R$. All other constants can be collected to form an effective ``Newtonian'' constant.

In this complementary picture, the prefactor of the interaction 
plays the role of an emergent source parameter, either ``charge-
like'' or ``mass-like'', depending on the chosen analogy. 
We stress that these emergent quantities are not fundamental 
charges or masses, but rather effective constructs that capture 
how collective excitations act as sources of long-range 
interactions mediated by the vacuum. 
This dual interpretation underscores the distinctive role of the MB-QED interaction, bridging familiar Coulomb and gravitational 
forms while remaining rooted in many-body QED.
An intriguing question is whether the effective mass and 
charge scales identified here can be related to the 
renormalization of mass and charge in interacting quantum field 
theories. In QED, it is 
well established that the particle mass and electric charge, 
together with the coupling constant, are renormalized by 
interactions, a process often 
described as the dressing of matter degrees of freedom by 
the EMF excitations. In close analogy, we conjecture that the effective charge and mass corrections obtained here can be viewed as a renormalization of the QDO parameters (polarizability and characteristic frequency) induced 
by the dressing of MBD polarization modes with radiative EMF 
excitations.

There is an alternative semi-classical interpretation of the inverse distance interactions arising from non-linear contributions to the polarizability of many interacting atoms~\cite{distasio2014many}. If molecular polarizability were just a sum of atomic polarizabilities, the $1/R$ QED interaction would vanish. The interelectronic coupling makes the polarizability of molecules, materials, and nanostructures highly nonlinear~\cite{ambrosetti2016wavelike}. Because of this, we expect the inverse distance interaction to be quite general and scale up when complex materials interact. In this context, we conjecture that the inverse-distance interaction can be measured. The force between two modes can be written in atomic units as 
$F(R)\;\approx\;\frac{2\,\alpha_{\mathrm{fsc}}^{3}}{\pi}\,
\frac{\mu_1\mu_2\,E_1E_2}{R^{2}}\;\frac{E_h}{a_0}$,
where $\mu$ is the dipole matrix element (a.u.) and $E$ the excitation energy (Ha). The dipole moment is connected to the static polarizability through the standard oscillator-strength relation 
$\alpha_0\simeq 2|d|^2/E$, with $\mu \equiv |d|$. 
For an argon dimer ($A_0\simeq 11.1$, $E\simeq 0.07$ Ha) this yields forces of order $10^{-22}$--$10^{-20}$~N at $R=10$--$100$~nm, in the zepto-Newton range. For realistic collective excitations with $\mu\simeq 50$--$100$~D and $E\simeq 1$--$2$~eV, the scaling $F\propto \mu_{\mathrm{mode}}^{2}E^{2}/R^{2}$ boosts the signal to $10^{-21}$--$10^{-19}$~N at separations of $50$--$200$~nm. These magnitudes directly overlap with the sensitivity of precision Casimir measurements at micro- and nanometric separations---using torsional oscillators, microcantilevers, or levitated particles---which already operate in the atto- to zepto-Newton regime~\cite{geraci2010short,sushkov2011new}. Likewise, short-range gravity tests with microfabricated resonators and optomechanical sensors are now entering the sub-100~$\mu$m domain~\cite{PhysRevD.110.122005}, where Casimir backgrounds dominate the signal budget; the distinct $1/R$ scaling provides a clear discriminant from both Casimir ($\sim 1/R^7$--$1/R^8$) and Newtonian ($\sim 1/R^2$) forces which are searched for in the experiments. The MB-QED contribution gives rise to a force scaling as $1/R^{2}$, placing it within the detection limit of the present Casimir and short-range gravity experiments (atto- to zepto-Newton).


Recent proposals \cite{berezhiani2019emergent,o2000bose} have shown that Bose–Einstein condensates can host emergent long-range forces with $1/R$ character, either mediated by electromagnetic fields or arising from collective modes. Our observation of a repulsive $1/R$ potential provides a complementary scenario, suggesting that such effective interactions could compete with or counterbalance the attractive mechanisms discussed in Bose-Einstein condensation contexts. Exploring this interplay experimentally would clarify whether condensates can support stable states governed by a balance of attractive and repulsive long-range forces.

The present approach is based on the QDO model for matter and on a second-order perturbative expansion in the field--matter interaction. As a result, the renormalized masses and self-energies that we derive should be interpreted as effective, model-dependent parameters rather than as universal renormalization constants of QED. In particular, their value depends on the chosen cutoff and on the QDO parametrization, and they cannot be directly identified with the electron mass renormalization of full QED. Furthermore, the restriction to second order means that the validity of our results is limited to the weak-coupling regime, where higher-order photon processes and collective dressing effects can be safely neglected. A natural next step is to pursue a fully non-perturbative treatment of the coupled matter--field system. This would allow one to include photon dressing and mode mixing self-consistently at all orders, without relying on cutoff regularization. Possible avenues include functional-integral techniques, variational approaches to the full field--matter Hamiltonian, or diagrammatic resummations tailored to collective polarization modes. Such methods would clarify the robustness of the $1/R$ term, provide controlled predictions beyond weak coupling, and establish closer connections with renormalization theory in quantum electrodynamics.

We warmly thank Akbar Salam, Ulf Leonhardt, and Roberto Passante for carefully reading the manuscript and for their valuable comments and constructive suggestions.
 


\clearpage
\onecolumngrid
\begin{center}\textbf{\Large Supplemental Material\\
Repulsive Inverse-Distance Interatomic Interaction from \\Many-Body Quantum Electrodynamics}\end{center}

\setcounter{equation}{0}\setcounter{figure}{0}\setcounter{table}{0}
\renewcommand{\theequation}{S\arabic{equation}}
\renewcommand{\thefigure}{S\arabic{figure}}
\renewcommand{\thetable}{S\arabic{table}}

\section{QDO model for matter system}

The quantum Drude oscillator (QDO) is a coarse-grained quantum-mechanical model that describes the electronic response of valence electrons in atoms and molecules~\cite{Jones2013}. The model is composed of two pseudo-charges, an infinitely heavy nucleus(--like) positive charge $q_{nucl}=-q$ located at the point $\bm{R}$, and a negatively charged Drudon $q $ with the position vector $\bm{x}$, bound to the nucleus via a harmonic potential,
\begin{equation}
\frac{1}{2}m\omega^2|\bm{x}-\bm{R}|^2\ .
\end{equation}
The model parameters, $m$ and $\omega$, respectively, denote the mass and the frequency of the QDO.
The interaction of the Drude oscillator with the (vacuum) electromagnetic field can be introduced through the minimal coupling principle. However, we first note that we are interested in long-wave phenomena related to dipolar properties of the matter system. Therefore, we assume that the vector potential does not appreciably vary on the characteristic length scale of the matter system. In the other word, we assume that $\bm{A}(\bm{x})\approx \bm{A}(\bm{R})$, hence the minimal coupling is given by
\begin{equation}\label{def:minimal_coupling}
    \bm{p}\mapsto \bm{p} - q\bm{A}(\bm{R})\ ,
\end{equation}
where $\bm{p}$ is the momentum of the Drudon. 
Thus, the Hamiltonian of the matter system (modeled 
by a single QDO) interacting with the free 
electromagnetic field is given by
\begin{equation}
\label{SM_eq:total-Hamiltonian}
   H = H_{\rm matter} + H_{\rm field} + H_{\rm int} = \frac{1}{2m}\left|\bm{p} - q\bm{A}(\bm{R})\right|^2+\frac{1}{2}m\omega^2|\bm{x}-\bm{R}|^2 + H_{\rm field}\ .
\end{equation}
The first term in the Hamiltonian produces three terms (in the Coulomb gauge) that read
\begin{equation}
    \frac{1}{2m}\left|\bm{p} - q\bm{A}(\bm{R})\right|^2 = 
    \frac{1}{2m}\bm{p}^2- \frac{q}{m}\bm{p}\cdot \bm{A}(\bm{R})
    + \frac{q^2}{2m}\bm{A}^2 \ ,
\end{equation}
so that we recognize the interaction terms to be
\begin{equation}
    H_{\rm int}=- \frac{q}{m}\bm{p}\cdot\bm{A}(\bm{R})+ \frac{q^2}{2m}\bm{A}^2(\bm{R})\ .
\end{equation}
The quantization of the QDO Hamiltonian is done
by following the Dirac's prescription: we promote $(\bm{p},\bm{x})$ to quantum mechanical operators satisfying the canonical commutation relation
\begin{equation}
    [\hat{x}_{ia} , \hat{p}_{jb}]=i\delta_{ab}\delta_{ij}\hbar .
\end{equation}
where the indexes $a,b$ labels respectively the
$a$-th and $b$-th atomic QDOs and the indexes $i,j=x,y,z$ labels the 
cartesian coordinates.

The matter--matter interaction for two QDOs is the Coulomb coupling between the charged (pseudo) particles of the two entities. In the dipole approximation, this QDO--QDO interaction is given by 
\begin{equation}
\label{SM_eq:V_CC-1}
    \hat{V}_C(\hat{\bm{x}}_1,\hat{\bm{x}}_2) 
    \approx \hat{V}_{\rm DD}(\hat{\bm{r}}_1, \hat{\bm{r}}_2, \bm{R}) = \frac{q_1q_2}{4\pi\epsilon_0\, R^3} 
    \left[\hat{\bm{r}}_1\cdot\hat{\bm{r}}_2 - 3\dfrac{(\bm{R}\cdot \hat{\bm{r}}_1) (\bm{R}\cdot \hat{\bm{r}}_2)}{R^2} \right] \ ,
\end{equation}
where $\hat{\bm{r}}_a$ is the $a$-th atomic Drudon position relative to the atomic center, i.e. $\hat{\bm{r}}_a=\hat{\bm{x}}_a-\bm{R}_a$, and $\bm{R}=\bm{R}-\bm{R}_1$ is the relative position between the atomic nuclei. 
For simplicity, we assume that $\bm{R}$ is along the z-axis, so  $\bm{R}=R\hat{e}_z$ being $\hat{\boldsymbol{e}}_z$ the unity vector along the z-axis.
Therefore, the matter Hamiltonian, including the dipole–dipole Coulomb interaction 
$\hat{V}^{\rm dip-dip}_{12}$, can be written as  
\begin{equation}
\label{SM_eq:H_matter}
\hat{H}_{\rm matter} = 
\sum_{a=1,2} \left[ \frac{\hat{\bm{p}}_a^2}{2m_a} + \frac{1}{2} m_a \omega_a^2 \hat{\bm{r}}_a^2 \right] 
+ \frac{q_1 q_2}{4\pi \epsilon_0 R^3} 
\left( \hat{\bm{r}}_1 \cdot \hat{\bm{r}}_2 - 3 \hat{z}_1 \hat{z}_2 \right)
= \sum_{a=1,2} \frac{\hat{\bm{p}}_a^2}{2m_a} 
+ (\hat{\bm{r}}_1, \hat{\bm{r}}_2)\, \mathbb{V}\, 
\begin{pmatrix}
\hat{\bm{r}}_1 \\ \hat{\bm{r}}_2
\end{pmatrix},
\end{equation}
where we have introduced the potential energy matrix
\begin{equation}
\mathbb{V} =
\begin{pmatrix}
\omega_1^2 \mathbb{I}_3 & \mathbb{T}(\boldsymbol{R}_{12}) \\[6pt]
\mathbb{T}(\boldsymbol{R}_{12}) & \omega_2^2 \mathbb{I}_3
\end{pmatrix},
\end{equation}
with $\mathbb{T}(\boldsymbol{R}_{12})$ denoting the dipole–dipole coupling matrix.

\section{Diagonalization of the matter Hamiltonian}
In the case of a quadratically coupled quantum harmonic oscillator, such as for $\hat{H}_{\rm matter}$, the reduction of the Hamiltonian to its normal form is equivalent to the diagonalization of the symmetric, positive-definite matrix associated with the quadratic potential.
We define the set of mass scaled coordinates 
$\hat{\boldsymbol{\xi}}_a = m^{1/2} \hat{\boldsymbol{r}}_a$ and 
momenta $\hat{\bm{\pi}}_a=m_a^{-1/2}\hat{\bm{p}}_a$. 
We notice here that the mass scaling transformation reported above
preserve the canonical commutation relations (CCR), i.e. $[\hat{\xi}_{ia},\hat{\pi}_{jb}]= i\hbar \delta_{ij}\delta_{ab}$ and $[\hat{\xi}_{ia},\hat{\xi}_{jb}]=[\hat{\pi}_{ia},\hat{\pi}_{jb}]=0$.
Moreover, by introducing the coupling parameters
\begin{equation}
\label{eq:MBD_gamma_strength}
\gamma_x=\gamma_y=\gamma=\frac{q_1q_2}{4\pi\epsilon_0 \sqrt{m_1m_2} R^3} 
\ , \quad \text{and} \quad \gamma_z=-2\gamma
\end{equation}
it is possible to rewrite the matter Hamiltonian $\hat{H}_{\rm matter}$ as follows
\begin{equation}
\label{SM_eq:H_matter_2}
\hat{H}_{\rm matter} = \sum_{i=x,y,z} \left[ \sum_{a=1,2} \left(
\frac{\hat{\pi}_{ia}^2}{2} + \frac{1}{2}\omega_a^2\hat{\xi}_{ia}^2\right)
+ \gamma_i\hat{\xi}_{i1}\hat{\xi}_{i2} \right] \ .
\end{equation}
In matrix representation, it is straightforward to see that  Hamiltonian~\eqref{SM_eq:H_matter_2} has the eigenvalues 
\begin{equation}
\label{SM_eq:matter_eigenvalues}
\lambda_{i,\pm} = \frac{1}{4}\left[\omega_1^2+\omega_2^2 \pm \sqrt{4\gamma_i^2 +(\omega_1^2-\omega_2^2)^2} \right] 
\ , \  i=x,y,z \ ,
\end{equation} 
with the corresponding normalized eigenvectors
\begin{equation}
\label{SM_eq:matter_eigenvectors}
v_{i,\pm} = \left[\frac{\gamma_i^2}{4\gamma_i^2 + [\omega_1^2-\omega_2^2 \pm \sqrt{4\gamma_i^2+(\omega_1^2-\omega_2^2)^2}\,]^2}\right]^\frac{1}{2}
\begin{pmatrix}
\frac{1}{\gamma_i}[\omega_1^2-\omega_2^2 \pm \sqrt{4\gamma_i^2+(\omega_1^2-\omega_2^2)^2}~]
\\

\\
2
\end{pmatrix}
\ , \  i=x,y,z \ .
\end{equation}

The diagonalization of the matrix $\mathbb{V}$ 
allows to define a set of normal modes coordinates $\hat{\xi}_{i\pm}$
\begin{gather}
\label{SM_eq:normal_mode_coordinates}
\zeta_{i+} = \mathcal{N}_{i+}^{(1)}\xi_{i,1} + \mathcal{N}_{i+}^{(2)}\xi_{i,2} \ ,
\quad 
\zeta_{i-} = \mathcal{N}_{i-}^{(1)}\xi_{i,1} + \mathcal{N}_{i-}^{(2)}\xi_{i,2} \ ,
\end{gather}
with 
\begin{gather}
\label{SM_eq:Omat}
\mathcal{N}_{i\pm}^{(1)} = \frac{\mathcal{B}_{i\pm} }{\gamma_i}\left[\omega_1^2-\omega_2^2 \pm \sqrt{4\gamma_i^2+(\omega_1^2-\omega_2^2)^2}\right] \ ,
\quad \quad \text{and} \quad \quad
\mathcal{N}_{i\pm}^{(2)} = 2\mathcal{B}_{i\pm} \ ,
\\
\mathcal{B}_{i\pm} = \left[\frac{\gamma_i^2}{8\gamma_i^2 + 2 (\omega_1^2-\omega_2^2) [\omega_1^2-\omega_2^2 \pm \sqrt{4\gamma_i^2+(\omega_1^2-\omega_2^2)^2}\,]^2}\right]^\frac{1}{2} \ .
\end{gather}
In a compact, form we will rewrite the previous change of variables as $\boldsymbol{\zeta} = \mathcal{N} \boldsymbol{\xi}$. The corresponding normal-mode momenta $\eta_{i\pm}$ are defined such that 
$\eta_{i,a} = \mathcal{N}_{i+}^{(a)}\pi_{i+} + \mathcal{N}_{i-}^{(a)}\pi_{i-}$ (with $a=1,2$ and $i=x,y,z$). 
The normal-mode coordinates $\xi_{i\pm}$ and momenta $\pi_{i\pm}$ reduce to the simple expressions $\zeta_{i\pm} = \frac{1}{\sqrt{2}}(\xi_{i,2}\pm\xi_{i,1})$ and $\eta_{i\pm} = \frac{1}{\sqrt{2}}(\pi_{i,2}\pm\pi_{i,1})$ if the original isolated QDOs are resonant. 
The normal-mode coordinates and momenta can be considered as those of 
six one-dimensional normal-mode oscillators with characteristic 
frequencies $\omega_{i\pm}^2 =2 \lambda_{i,\pm}$ for $i=x,y,z$. 
The matter Hamiltonian in Eq.\eqref{SM_eq:H_matter_2}  can be rewritten in the normal form
\begin{gather}
\label{SM_eq:H_matter_3}
\hat{H}_{\rm matter} = \sum_{\alpha=\pm}\ \sum_{i=x,y,z} \dfrac{\hat{\eta}_{i\alpha}^2}{2}+\dfrac{1}{2}\omega_{i\alpha}^2\hat{\zeta}_{i\alpha}^2 \ ,
\end{gather}

It is convenient, for further computations, to express such a diagonalization 
procedure of the MBD Hamiltonian in terms of the CCR-preserving transformation
from the creation/annihilation operator algebra of the atomic QDOs 
\begin{equation}
    \hat{\xi}_{ia} = \sqrt{\dfrac{\hbar}{2 \omega_a}} (\hat{a}_{ai}^{\dagger}+\hat{a}_{ai}) \qquad \hat{\pi}_{ia} = i\sqrt{\dfrac{\hbar \omega_a}{2}} (\hat{a}_{ai}^{\dagger}-\hat{a}_{ai})
\end{equation}
to the creation/annihilation operator algebra of the MBD normal modes
\begin{equation}
    \hat{\zeta}_{i\alpha} = \sqrt{\dfrac{\hbar}{2 \omega_{i\alpha}}} (\hat{b}_{i\alpha }^{\dagger}+\hat{b}_{ i\alpha }) \qquad \hat{\eta}_{i\alpha} = i\sqrt{\dfrac{\hbar  \omega_{i\alpha}}{2}} (\hat{b}_{i\alpha}^{\dagger}-\hat{b}_{i\alpha})\,\,.
\end{equation}
Following the results derived in \cite{gori2023second} the Bogoliubov
transformation between creation/annihilation operator of atomic modes and 
MBD modes
\begin{equation}
\label{eq:Bogo_trans}
    \begin{pmatrix}
    \hat{\boldsymbol{b}}\\
    \hat{\boldsymbol{b}}^{\dagger}\\
    \end{pmatrix}
    =
    \begin{pmatrix}
    X & Y \\
    Y & X
    \end{pmatrix}
        \begin{pmatrix}
    \hat{\boldsymbol{a}}\\
    \hat{\boldsymbol{a}}^{\dagger}\\
    \end{pmatrix}
\end{equation}
where the $X=X(\mathcal{N},\boldsymbol{\omega},\tilde{\boldsymbol{\omega}})$ and $Y=Y(\mathcal{N},\boldsymbol{\omega},\tilde{\boldsymbol{\omega}})$  are the $3N \times 3N$ defined as
\begin{equation}
     X = \dfrac{1}{2}\left[\widetilde{\Omega}^{1/2}\mathcal{N}\Omega^{-1/2} + \widetilde{\Omega}^{-1/2}\mathcal{N}\Omega^{1/2}\right]\qquad  Y = \dfrac{1}{2}\left[\widetilde{\Omega}^{1/2}\mathcal{N}\Omega^{-1/2} - \widetilde{\Omega}^{-1/2}\mathcal{N}\Omega^{1/2}\right]
\end{equation}
where $\widetilde{\Omega}=\mathrm{diag}\{\omega_{x,+},\omega_{x,-},....\}$ is the diagonal matrix of normal modes eigenfrequencies and
$\Omega=\mathrm{diag}\{\omega_{x,a},\omega_{x,a},....\}$ is the diagonal matrix of atomic QDOs eigenfrequencies.

These matrices allows to define the transformation for the non-dimensionalized quantities, i.e.
\begin{equation}
  \hat{\xi}_{ia}= \sqrt{\dfrac{\hbar}{2\omega_{ia}}} (\hat{a}^{\dagger}_{ia}+\hat{a}_{ia})= \sum_{\alpha=\pm}  \sqrt{\dfrac{\hbar}{2\omega_{ia}} }\mathcal{M}^{(a)}_{i\alpha}(\hat{b}^{\dagger}_{i\alpha}+\hat{b}_{i\alpha}) \qquad \hat{\pi}_{a_i} = i\sqrt{\dfrac{\hbar \omega_{ia}}{2}} (\hat{a}^{\dagger}_{a_i}-\hat{a}_{a_i})= i\sqrt{\dfrac{\hbar \omega_{ia}}{2}}  \sum_{\alpha=\pm} \mathcal{S}^{(a)}_{i\alpha}(\hat{b}^{\dagger}_{\alpha} - \hat{b}_{\alpha})
\end{equation}
where $\mathcal{M}^{(a)}=X^{T} - Y^{T}$ and $\mathcal{S}=X^{T} + Y^{T}$.
The eigenstates of the MBD Hamiltonian constitute the reference basis set to express 
the matter state and they are expressed as
\begin{align}
\label{SM_eq:eigenvalues_eigenstates_matter}
& |\boldsymbol{n}\rangle = \bigotimes_{i=1,2,3}\bigotimes_{\alpha=\pm} |n_{i\alpha} \rangle \qquad  E_{\rm matter} = \sum_{\alpha=\pm}\ \sum_{i=x,y,z}\,\,\hbar\omega_{i \alpha}\left(n_{i \alpha}+\frac{1}{2}\right)\,\,.
\end{align}
where $\boldsymbol{n}\in\mathbb{N}^{6}$ is the label of MBD eigenstate and $n_{i \alpha}\in\mathbb{N}$ is the occupation number of the MBD model labeled by the indexes $(i,\alpha)$.

\section{Quantum electromagnetic field}

The free electromagnetic field is quantized in a cubic box of volume 
$V$, providing at the same time the boundary conditions for the 
Maxwell equations and the infrared regularization. 
Within the box a natural basis set is constituted by plane waves
with wavevector $\bm{k}$. The polarizations are determined by the  unit vectors $\boldsymbol{e}_{\lambda}(\boldsymbol{k})$ with $\lambda=1,2$ satisfying
the following completeness relation
\begin{equation}
\label{eq:completness_pol}
    \sum_{\lambda=1,2} e_i(\boldsymbol{k},\lambda) e_j(\boldsymbol{k},\lambda) =\delta_{ij} - \hat{k}_{i}\hat{k}_{j}\,\,.
\end{equation}
Such a physical picture is equivalent to the Fourier expansion of the free 
field in terms of the plane waves solving the source-free Maxwell equations.
Hence, the field $\bm{A}$ is quantized in the same way an oscillator is,
{\it i.e.} promoting the Fourier expansion coefficients of each mode to quantum 
mechanical operators that obey canonical commutation relations
\begin{equation}
[\hat{a}_{\lambda}(\bm{k}), \hat{a}_{\lambda'}^\dagger(\bm{k}')] = \delta_{\bm{k}\bm{k}'}\delta_{\lambda\lambda'} \qquad [\hat{a}_{\lambda}(\bm{k}), \hat{a}_{\lambda'}(\bm{k}')] = 
[\hat{a}_{\lambda}^\dagger(\bm{k}), \hat{a}_{\lambda'}^\dagger(\bm{k}')] = 0
\end{equation}
with $\hat{a}_{\lambda}^\dagger(\bm{k})$ and $\hat{a}_{\lambda}(\bm{k})$ indicating 
creation and annihilation operators of the photon in the mode $(\bm{k},\lambda)$. 
Therefore, the quantized field $\bm{A}$ is given by 
\begin{equation}
\label{def:vector_potential}
    \hat{\bm{A}}(\bm{x},t)=\sum_{\bm{k}\in\mathcal{S}(V,k_{\rm M})}\sum_{\lambda=1}^{2}\sqrt{\frac{\hbar}{2\epsilon_0 c k V}}  
    \left[\bm{e}(\bm{k},\lambda) \hat{a}_{\lambda}(\bm{k})e^{i\bm{k}\cdot\bm{x}} 
    + \bm{e}(\bm{k},\lambda) \hat{a}^{\dagger}_{\lambda}(\bm{k}) e^{-i\bm{k}\cdot\bm{x}}\right] \ ,
\end{equation}
where $\mathcal{S}(V,k_{\rm M})=\{\boldsymbol{k}\in\mathbb{R}^3 \,\,|\,\, \bm{\eta} \in \mathbb{Z} \,\,\wedge \,\, \bm{k}=2\pi \boldsymbol{\eta}/\sqrt[3]{V}\,\, \wedge\,\, |\bm{k}|\leq k_{M} \}$ is the integration domain.

By using the mode-decomposition of the vector field
presented in Eq.\eqref{def:vector_potential}, the
Hamiltonian of the free EM field can be rewritten as 
\begin{equation}
    \hat{H}_{\rm field}=\sum_{\boldsymbol{k}\in\mathcal{S}(V,k_{\rm M})}\sum_{\lambda=1}^2 \hbar ck \left(\hat{n}_{\boldsymbol{k},\lambda}+\dfrac{1}{2}\right)
\end{equation}
where $\hat{n}_{\boldsymbol{k},\lambda}$ is the number operators
of the photons in the mode $(\boldsymbol{k},\lambda)$.
Adding a second QDO to the matter system introduces new terms to the 
Hamiltonian~\eqref{SM_eq:total-Hamiltonian}, accounting for additional
matter--matter and matter--field interactions. 

Introduction of a second QDO does not change the Hamiltonian of the free field, but it 
adds two new matter--field terms that take into account the interaction of the second 
QDO with the vacuum field, hence 
\begin{equation}
\label{SM_eq:H_int}
\hat{H}_{\rm int} = \sum_{a=1,2} \left[
- \frac{q_a}{m_a}\hat{\bm{p}}_a\cdot\hat{\bm{A}}(\bm{R}_a) 
+ \frac{q_a^2} {2m_a}\hat{\bm{A}}^2(\bm{R}_a) \right]\ .
\end{equation}

\subsection*{Diamagnetic contribution in perturbation theory}

In the perturbative framework adopted below, we employ as a basis the tensor product between the Hilbert space of MBD modes and the Fock basis for the photon field. The unperturbed ground state is
\begin{equation}
    |GS\rangle = |0_{\text{MBD}}\rangle \otimes |0_{\text{phot}}\rangle,
\end{equation}
with
\[
    \widehat{a}_{\lambda}(\bm{k})|0_{\text{phot}}\rangle = 0, 
    \qquad 
    \widehat{b}_{ia}|0_{\text{MBD}}\rangle = 0.
\]

The diamagnetic term in the minimal-coupling Hamiltonian reads
\begin{equation}
    \widehat{H}^{(2)}_a = \frac{q_a^2}{2m_a}\,\widehat{\bm{A}}(\bm{R}_a)^2,
\end{equation}
where the vector potential operator in the Coulomb gauge is expanded as
\begin{equation}
    \widehat{A}_i(\bm{x})
    =
    \sum_{\bm{k},\lambda}
    \sqrt{\frac{\hbar}{2\varepsilon_0 c k V}}
    \,\epsilon^i_{\lambda}(\bm{k})
    \Big(\widehat{a}_{\lambda}(\bm{k})e^{i\bm{k}\cdot\bm{x}}
    +\widehat{a}_{\lambda}^{\dagger}(\bm{k})e^{-i\bm{k}\cdot\bm{x}}\Big).
\end{equation}

\paragraph{First-order contribution.}
The first-order correction to the ground-state energy is
\begin{equation}
    \Delta E^{(1)}_{\bm{A}^2}
    = \langle GS|\widehat{H}^{(2)}|GS\rangle
    = \sum_a \frac{q_a^2}{2m_a}\,
    \langle 0_{\text{phot}}|\widehat{\bm{A}}(\bm{R}_a)^2|0_{\text{phot}}\rangle.
\end{equation}
Expanding $\widehat{\bm{A}}(\bm{R}_a)^2=\sum_i\widehat{A}_i(\bm{R}_a)\widehat{A}_i(\bm{R}_a)$ and taking the vacuum expectation value, only the contractions 
$\langle 0|\widehat{a}_{\lambda}(\bm{k})\widehat{a}^{\dagger}_{\lambda'}(\bm{k}')|0\rangle
=\delta_{\bm{k}\bm{k}'}\delta_{\lambda\lambda'}$ survive. Thus,
\begin{equation}
    \langle 0_{\text{phot}}|\widehat{\bm{A}}(\bm{R}_a)^2|0_{\text{phot}}\rangle
    =
    \sum_{\bm{k},\lambda}
    \frac{\hbar}{2\varepsilon_0 c k V},
\end{equation}
and therefore
\begin{equation}
    \Delta E^{(1)}_{\bm{A}^2}
    =
    \sum_a\frac{q_a^2}{2m_a}
    \sum_{\bm{k},\lambda}
    \frac{\hbar}{2\varepsilon_0 c k V}.
\end{equation}
This contribution depends only on local quantities ($q_a$, $m_a$) and is independent of the interatomic distance $R$. It represents a purely local self-energy shift, and thus it does not contribute to the interaction energy between QDOs.

\paragraph{Second-order contribution.}
At second order, the diamagnetic term yields
\begin{equation}\label{def:second-order-A2}
    \Delta E^{(2)}_{\bm{A}^2}
    =
    -\sum_{n\neq GS}
    \frac{|\langle n|\widehat{H}^{(2)}|GS\rangle|^2}{E_n-E_{GS}},
\end{equation}
where $|n\rangle$ runs over all excited photon states of the Fock space,
\[
    |n\rangle
    =|0_{\text{MBD}}\rangle
    \otimes
    \sum_{\bm{k},\lambda} |n_{\bm{k}\lambda}\rangle_{\text{phot}}.
\]
In principle, all multiphoton states $|2_{\bm{k}\lambda};2_{\bm{k}'\lambda'}\rangle$, $|3_{\bm{k}\lambda};3_{\bm{k}'\lambda'}\rangle$, etc.\ should be included.
However, to identify the dependence on the charge $q$, it suffices to consider the lowest-order non-vanishing process, in which $\widehat{\bm{A}}^2$ creates two photons from the vacuum through the term $\widehat{a}^{\dagger}\widehat{a}^{\dagger}$. The relevant matrix element is therefore
\begin{equation}
    \langle n|\widehat{H}^{(2)}|GS\rangle
    \sim
    \sum_a\frac{q_a^2}{2m_a}\,
    \langle 2_{\bm{k}\lambda};2_{\bm{k}'\lambda'}|
    \widehat{\bm{A}}(\bm{R}_a)^2
    |0_{\text{phot}}\rangle.
\end{equation}
From the mode expansion, this element is proportional to
\begin{equation}
    \langle n|\widehat{H}^{(2)}|GS\rangle
    \propto
    \frac{q_a^2}{m_a}
    \frac{\hbar}{\varepsilon_0 c\sqrt{k k'}V}.
\end{equation}
The corresponding squared amplitude scales as
\begin{equation}
    |\langle n|\widehat{H}^{(2)}|GS\rangle|^2
    \propto
    \left(\frac{q_a^2}{m_a}\right)^2
    \left(\frac{\hbar}{\varepsilon_0 c}\right)^2
    \frac{1}{k k'V^2}.
\end{equation}
Since the energy denominator in Eq.~\eqref{def:second-order-A2} scales as $E_n-E_{GS}\sim\hbar c(k+k')$, we obtain the scaling behavior
\begin{equation}
    \Delta E^{(2)}_{\bm{A}^2}
    \propto
    \left(\frac{q_a^2}{m_a}\right)^2
    \sim O(q^4).
\end{equation}

In conclusion, the $\widehat{\bm{A}}^2$ term contributes at first order with a local, distance-independent energy shift, and at second order with terms of order $O(q^4)$.  
Both effects are of higher order compared to the leading $O(q^2)$ correction arising from the $\widehat{\bm{p}}_a\!\cdot\!\widehat{\bm{A}}$ coupling.  
Therefore, within the present perturbative framework, the $\widehat{\bm{A}}^2$ term can be safely neglected when evaluating the interaction energy between QDOs. The extended calculation of $\Delta E^{(2)}_{\bm{A}^2}$ will be performed in Sec.~\ref{sec:energy-shift-A2}.

\section{Second-order perturbation to the matter--field system}

We use second-order perturbation theory to 
obtain the energy shift due to the QDOs--field 
interactions.
First, we consider the first term of $H_{\rm int}$ that is linear 
in $\bm{A}$, and rewrite it in terms of the normal-mode momentum operators,  
\begin{align}
\hat{H}_{\rm int}^{(1)} &= - \sum_{a=1,2}
 \frac{q_a}{m_a}\hat{\bm{p}}_a\cdot\hat{\bm{A}}(\bm{R}_a)=
-\sum_{a=1,2}\sum_{i=x,y,z} 
\frac{q_a}{\sqrt{m_a}} \hat{\pi}_{ia}\, \hat{A}_i(\bm{R}_a) =
\nonumber\\
&=
- i\sum_{a=1,2}\sum_{\alpha=\pm} \sum_{i=x,y,z} 
\sum_{\boldsymbol{k}\in \mathcal{S}(V,k_{\rm M})}\sum_{\lambda=1}^{2}\sqrt{\frac{\hbar}{2\epsilon_0 c k V}}
\frac{q_a}{\sqrt{m_a}}\sqrt{\dfrac{\hbar\omega_a}{2}}\mathcal{M}_{i\alpha}^{(a)}\,\,(\hat{b}^{\dagger}_{ia}-\hat{b}_{ia}) \\
&
\times \left[e_i(\bm{k},\lambda) \hat{a}_{\lambda}(\bm{k})e^{i\bm{k}\cdot\bm{R}_a} 
    + e_i^*(\bm{k},\lambda) \hat{a}^{\dagger}_{\lambda}(\bm{k}) e^{-i\bm{k}\cdot\bm{R}_a}\right]\,\, .
\label{H_int(1)_+-}
\end{align}
The first-order correction to ground state energy are null as 
\begin{equation}
    \langle 0;\boldsymbol{0}_{\boldsymbol{k},\lambda} | \hat{H}^{(1)}_{\rm int} | 0;\boldsymbol{0}_{\boldsymbol{k},\lambda} \rangle = 0
\end{equation}
due to selection rules, i.e. $\langle 0| \hat{\eta}_{i\alpha}|0\rangle = 0$ and  $\langle \boldsymbol{0}_{\boldsymbol{k},\lambda}| \hat{a}_{\lambda}(\boldsymbol{k})|\boldsymbol{0}_{\boldsymbol{k},\lambda}\rangle= \langle \boldsymbol{0}_{\boldsymbol{k},\lambda}| \hat{a}^{\dagger}_{\lambda}(\boldsymbol{k})|\boldsymbol{0}_{\boldsymbol{k},\lambda}\rangle = 0$.
Therefore, the first non-zero correction is expected to be at the second order
\begin{gather}
\label{2ndpert-1}
\Delta E^{(pA,2)}= \sum_{I\neq 0}\frac{\langle 0|\hat{H}_{\rm int}^{(1)}|I\rangle \langle I|\hat{H}_{\rm int}^{(1)}|0\rangle}{E_0-E_I}\ ,
\end{gather}
where, due to selection rules, $|I\rangle = |1_{i\alpha};\boldsymbol{1}_{\boldsymbol{k},\lambda}\rangle$ is a state with a single exciton in the collective MBD mode $(i\alpha)$.
By using the second quantized formalism for the matter system to compute the expectation values, one obtains
\begin{equation}
\langle 0|i(\hat{b}^{\dagger}_{i\alpha}-\hat{b}_{i\alpha})|1_{i\beta}\rangle \langle 1_{i\beta}|i(\hat{b}^{\dagger}_{i\alpha}-\hat{b}_{i\alpha})|0\rangle = \|\langle 0|i(\hat{b}^{\dagger}_{i\alpha}-\hat{b}_{i\alpha})|1_{i\beta}\rangle \|^2 = \delta_{\alpha\beta}
\end{equation}
so that after a straightforward computation 
\begin{align}
&\Delta E^{(pA,2)} = -\sum_{a,a'=1}^2\, \sum_{\alpha=\pm}\, \sum_{i=x,y,z}\, 
\sum_{\bm{k}\in \mathcal{S}(V,k_{\rm M})}\sum_{\lambda=1}^{2}\left(\frac{\hbar}{2\epsilon_0 c k V}\right)
\left(\frac{q_j q_{j'}}{\sqrt{m_j m_{j'}}}\right)\mathcal{M}_{i\alpha}^{(a)} \mathcal{M}_{i\alpha}^{(a')}
\left(\frac{\hbar \sqrt{\omega_{ia}\omega_{ia'}}}{2}\right) 
\,\\
\nonumber &\times e_i(\bm{k},\lambda) e_i^*(\bm{k},\lambda)
\frac{e^{i\bm{k}\cdot (\bm{R}_j - \bm{R}_{j'})}}{\hbar (\omega_{k\lambda} + \omega_{i\zeta})}\ .
\end{align}
To simplify we introduce the normalized QDOs (self-)correlation $\mathcal{C}^{(aa')}_{i\alpha}= \mathcal{M}^{(a)}_{i\alpha}\mathcal{M}^{(a')}_{i\alpha}$ mediated by the MBD mode $(i,\alpha)$ and the non-dimensionalized electric charges
$Z_{j}=q_j/e$. Moreover, we will consider the continuous limit for the EM field
\begin{equation}
\dfrac{1}{V}\sum_k \longrightarrow \int \dfrac{d^3k}{8\pi^3}
\end{equation},
and after transforming $k$-integral into spherical coordinates and using the 
completeness relation for the polarizations $\boldsymbol{e}(\boldsymbol{k},\lambda)$ one obtain
\begin{equation}
\Delta E^{(pA,2)} = -\alpha_{\rm fsc}\sum_{a,a'=1}^2\,\sum_{\alpha=\pm}\,\sum_{i=x,y,z} 
\frac{\hbar^2}{8 \pi^2} 
\dfrac{ Z_a Z_{a'}}{\sqrt{m_a m_{a'}}}\mathcal{C}^{(aa')}_{i\alpha}
\int_0^{k_M} \frac{\sqrt{\omega_{a}\omega_{a'}}\, k\, dk}{ck + \omega_{i\alpha}}
\int_{\mathbb{S}^2} (1-\hat{k}_i \hat{k}_i) e^{i\bm{k}\cdot\bm{R}_{aa'}} d\Omega \ . 
\label{eq:deltaE_1-nn'}
\end{equation}
where $\alpha_{\rm fsc}= e^2/(4\pi\epsilon_{0}\hbar c)$ 
is the fine structure constant and $\mathbb{S}^2$ is the 
(unitary) 2-sphere.\\
The first sum in expression Eq.\eqref{eq:deltaE_1-nn'} can 
be split in two terms.
One can be interpreted as a self-energy of the single 
atomic QDOs corrections $j=j'$
\begin{equation}
\Delta U^{\rm self}_{a} = -\alpha_{\rm fsc}\,\sum_{\alpha=\pm}\,\sum_{i=x,y,z} 
\frac{E_{a}}{8 \pi^2} 
\dfrac{ Z_a^2}{m_a c^2}\mathcal{C}^{(aa)}_{i\alpha}
\int_0^{E_{M}} \frac{E \mathrm{d}E}{E + \hbar \omega_{i\alpha}}
\int_{\mathbb{S}^2} (1-\hat{k}_i \hat{k}_i)d\Omega \ . 
\end{equation}
where $E=\hbar c k$ and $E_{a}=\hbar \omega_a$
By using the completness relations for the polarization of the EM field
\begin{equation}
    \int_{\mathbb{S}^2} (\delta_{ij}-\hat{k}_i \hat{k}_j) \mathrm{d}\Omega = \dfrac{8\pi}{3} \delta_{ij}
\end{equation}
and after few algebraic manipulations the expression and using
Bethe's renormalization scheme one obtains 
\begin{equation}
\label{eq:Ua_pa2_step2}
    \Delta U_{a}^{\rm self} = \dfrac{\alpha_{\rm fsc}}{3\pi}\dfrac{Q_a^2 E_{a}}{ c^2}\sum_{\alpha=1}^{3N}\sum_{i=x,y,z}\mathcal{C}^{(aa)}_{i\alpha}(\hbar\omega_{i\alpha})\int_{0}^{E_M}\dfrac{\mathrm{d}E}{E+\hbar\omega_{i\alpha}}\,\,.
\end{equation}
where for convenience we have introduced the mass reduced charges $Q_{a}=Z_a/m_a^{1/2}$.

The other contribution, with $j \neq j'$,
corresponds to the corrections to interatomic interaction energy, i.e. inversion symmetry $\boldsymbol{k}\longrightarrow -\boldsymbol{k}$ 
the angular integral can be rewritten as 
\begin{equation}
  \int_{\mathbb{S}^2} (1-\hat{k}_i \hat{k}_i) e^{i\bm{k}\cdot\bm{R}_{aa'}} d\Omega = \dfrac{1}{2} \int_{\mathbb{S}^2} (1-\hat{k}_i \hat{k}_i) \left(e^{i\bm{k}\cdot\bm{R}_{aa'}} + e^{-i\bm{k}\cdot\bm{R}_{aa'}}\right)d\Omega = \int_{\mathbb{S}^2} (1-\hat{k}_i \hat{k}_i) \cos(\boldsymbol{k}\cdot\boldsymbol{R}_{aa'}) d\Omega 
\end{equation}
so that the full expression can be rewritten as 
\begin{equation}
    \Delta V^{\rm int}_{12} = -\alpha_{\rm fsc}\sum_{\alpha=\pm}\,\sum_{i=x,y,z} 
\frac{\hbar^2}{4 \pi^2} 
\dfrac{ Z_1 Z_{2} \sqrt{\omega_{i1}\omega_{i2}}}{\sqrt{m_1 m_{2}}c^2}\mathcal{C}^{(12)}_{i\alpha}
\int_0^{E_M} \frac{ E\, dE}{ck + \omega_{i\alpha}}
\int_{\mathbb{S}^2} (1-\hat{k}_i \hat{k}_i) \cos\left(\boldsymbol{k}\cdot\boldsymbol{R}_{12}\right) d\Omega \,\, . 
\end{equation}
Following our conventions on the geometry of the QDOs dimer, the angular integral reads
\begin{equation}
\int_{\mathbb{S}^2} (1-\hat{k}_i \hat{k}_i) \cos\left(\boldsymbol{k}\cdot\boldsymbol{R}_{aa'}\right) d\Omega = \dfrac{4\pi}{3}\left[f_{\parallel}(kR_{aa'}) \delta_{i,z} + f_{\rm \perp}(kR_{aa'}) (1-\delta_{i,z}) \right]
\end{equation}
where we have introduced the special functions
\begin{align}
\label{eq:special_funct_fperp}
f_{\perp}(x) &= \dfrac{x \cos(x) + (x^2 - 1)\sin(x)}{x^3} \\
\label{eq:special_funct_fparal}
f_{\parallel}(x) &= \dfrac{2\left[\sin(x) - x \cos(x)\right]}{x^3} \,,
\end{align}
Finally we obtain the expression 
\begin{equation}
\label{eq:Vab_pa2_step2}
    \Delta V_{12}^{\rm int } = - \dfrac{\alpha_{\rm fsc}}{\pi} \dfrac{Q_1 Q_2 \sqrt{E_1 E_{2}}}{c^2}  \sum_{\alpha=1}^{3N} \sum_{i=x,y,z} \mathcal{C}^{(12)}_{i\alpha}\int_{0}^{E_M} \dfrac{E}{E+\hbar\omega_{i\alpha}} \left[ f_{\parallel}(kR_{12}) \delta_{i,z} + (1-\delta_{i,z}) f_{\perp}(kR_{12})) \right] \mathrm{d}E 
\end{equation}

\section{Analysis of the homoatomic dimer}
For the further analysis, it is convenient to rexpress the integral in Eq.\eqref{eq:Vab_pa2_step2} as a non-dimensionalized integral
\begin{equation}
 \Delta V^{\rm int}_{12} = - \Delta V \sum_{\alpha=\pm}\sum_{i=x,y,z} \int_{0}^{1} \mathcal{C}^{(12)}_{i\alpha} \dfrac{\bar{k}}{\bar{k}+\bar{k}_{\alpha}} \mathcal{F}_{i}(\bar{k}\bar{R}) \mathrm{d}\bar{k}
\end{equation}
where 
\begin{equation}
   \mathcal{F}_{i}(x) = f_{\parallel}(x) \delta_{i,z} + f_{\perp}(x) \left(1-\delta_{i,z}\right)
\end{equation}
and $\bar{k}=k/k_{M}$, $\bar{R}=k_M R$, and $\bar{k}_{\alpha} = E_{\alpha}/(\hbar c k_M)$.
With such substitutions, the integral is adimensional, and the energy scale reads
\begin{align}
    \text{SI units} \qquad &\Delta V  = \alpha_{\rm fsc} \dfrac{E_M}{\pi}  \dfrac{Q_1 Q_2 (E_1 E_2)^{1/2}}{c^2}\\
    \text{natural units} \qquad &\Delta V = \alpha_{\rm fsc} \dfrac{E_M Q_1 Q_2 (E_1 E_2)^{1/2}}{\pi}\\
    \text{atomic units} \qquad &\Delta V  = \alpha^3_{\rm fsc} E_{h} \dfrac{[E_M]_{a.u.} [(\mathcal{A}_{0,1} \mathcal{A}_{0,2})^{1/2}]_{a.u.} [(E_1 E_2)^{3/2}]_{a.u.}}{\pi}
\end{align}
where $E_{M} = \hbar c k_{M}$we have used that in atomic units $[c]_{a.u.} = \alpha_{\rm fsc}^{-1}$, $Q_a = Z_{a}/m_a$ is the mass reduced charge, and we have introduced the Hartree energy scale $E_h$.\\
In this case the previous expression can be further simplified by introducing the characteristic QDO wavenumber $\bar{k}_{\alpha} = \left[\omega_{\alpha}/(c k_M)\right] \sigma_{\alpha}(\gamma)  = \bar{k}_{Q} \sigma_{\alpha}(\gamma)$ where we have introduced the parameters $\gamma = \mathcal{A}_{0}/R_{12}^3$ and $\tau=R_{12} k_M$, i.e.
\begin{equation}
\label{eq:adim_DeltaV_homoatm}
 \Delta V^{\rm int}_{12} = - \Delta V \sum_{\alpha=\pm}\sum_{i=x,y,z} \int_{0}^{1} \mathcal{C}^{(12)}_{i\alpha}(\gamma)  \mathcal{F}_{i}(\bar{k}\tau) \dfrac{\bar{k} \mathrm{d}\bar{k}}{\bar{k}+\bar{k}_{Q}(\eta) \sigma_{\alpha}(\gamma)} = - \Delta V  \int_{0}^{1} \mathcal{I}(\bar{k};\gamma,\tau,\eta)  \mathrm{d}\bar{k}\,\,.
\end{equation}
In order to provide an analytic approximation, we proceed by 
approximating the integrand in equation 
\eqref{eq:adim_DeltaV_homoatm}. The $\bar{k}_{Q}(\eta)= \omega_{a}/c $ where we used the previous
definition of $k_{M}$. For the Argon homodimer $$\bar{k}_{Q}(\eta)= \omega_{a}\lambda_{\rm QDO,max}/(\eta c)= 2.52 \times 10^{-3}/\eta\,\,.$$
The correlation matrix and the MBD eigenfrequencies reads as a function of the $\gamma$ parameter
\begin{align}
 & \mathcal{C}^{(12)}_{z\pm} =  \dfrac{1}{2} \sqrt{1\pm \gamma} \left[ \delta_{ab}  \mp  (1-\delta_{ab})  \right]\\
&\mathcal{C}^{(12)}_{i\pm}=  \dfrac{1}{2} \sqrt{1\pm \gamma} \left[ \delta_{ab}  \mp  (1-\delta_{ab})  \right] (1-\delta_{i,z})\\
&\omega_{z\pm} =\omega \sqrt{1\pm 2 \gamma}\\
&\tilde{\omega}_{i\pm} = \omega \sqrt{1\pm \gamma}\qquad i=x,y
\end{align}
The exact expression of the integrand reads
\begin{align}
&\mathcal{I}(\bar{k};\gamma,\tau,\eta) = \frac{1}{\tau^3 \bar{k}^2} \left\{
\right. 
\frac{
  \sqrt{\gamma + 1} \left[ \left( \tau^2 \bar{k}^2 - 1 \right) \sin(\tau \bar{k}) + \tau \bar{k} \cos(\tau \bar{k}) \right]
}{
  \sqrt{\gamma + 1} \, \bar{k}_Q(\eta) + \bar{k}
}
+ 
\frac{
  \sqrt{2\gamma + 1} \left[ \tau \bar{k} \cos(\tau \bar{k}) - \sin(\tau \bar{k}) \right]
}{
  \sqrt{2\gamma + 1} \, \bar{k}_Q(\eta) + \bar{k}
}
\\[1em]
& +
\frac{
  \sqrt{1 - 2\gamma} \left[ \sin(\tau \bar{k}) - \tau \bar{k} \cos(\tau \bar{k}) \right]
}{
  \sqrt{1 - 2\gamma} \, \bar{k}_Q(\eta) + \bar{k}
}
+ 
\frac{
  \sqrt{1 - \gamma} \left[ \sin(\tau \bar{k}) - \tau \bar{k} \left( \tau \bar{k} \sin(\tau \bar{k}) + \cos(\tau \bar{k}) \right) \right]
}{
  \sqrt{1 - \gamma} \, \bar{k}_Q(\eta) + \bar{k}
}
\left. \right\}\,\,\,.
\end{align}
The full expression derived above is rather cumbersome, and obtaining a 
more transparent analytical form as a function of the intermolecular 
distance requires a series of simplifications. To this end, we first 
observe that for the Argon homodimer, the parameter \(\gamma\) can be 
approximated as \(\gamma = 11.1 / [R_{12}]^3_{\mathrm{a.u.}}\). This 
implies that in the short-range (strongly interacting) regime, one has \(\gamma|_{R_{12} = 5\,\text{\AA}} \approx 1.3 \times 10^{-2}\), whereas 
in the long-range (weakly interacting) regime, \(\gamma|_{R_{12} = 100\,\text{\AA}} \approx 1.6 \times 10^{-6}\). These values justify a 
perturbative treatment based on an expansion in small \(\gamma\).
However, a major challenge remains due to the presence of trigonometric 
functions of the variable \(\bar{k}\) in the integrand, which complicates 
analytical handling. A potential reduction in complexity can be achieved by performing a Taylor expansion in the limit \(\tau \rightarrow 0\). However, we verify that at large interatomic distances, specifically for \(R_{12} = 100\,\text{\AA}\) and \(\eta = 0.02\), the parameter \(\tau\) assumes the value \(\tau \sim 1.9 > 1\). While this does not entirely rule out the use of a Taylor expansion around \(\tau = 0\), it indicates that a more careful analysis is required to assess the convergence and accuracy of such an approximation.
For this aim, we introduce the approximate expression of the integrand:
\[
\mathcal{I}^{(l_1,l_2)}_{\text{appr}}(\bar{k}; \gamma, \tau, \eta) = \mathcal{I}(\bar{k}; 0, 0, \eta) + \sum_{l_1=1}^{m_\gamma} \sum_{l_2=1}^{m_\tau} \left. \partial_\gamma^{l_1} \partial_\tau^{l_2} \mathcal{I}(\bar{k}; \gamma, \tau, \eta) \right|_{\gamma=0,\, \tau=0} \, \gamma^{l_1} \tau^{l_2}.
\]
As a metric for the quality of the approximation, we consider
\begin{equation}
\frac{\left| \delta \mathcal{I}(\bar{k}; \gamma(R_{12}), \tau(R_{12}), \eta) \right|}
     {\left| \int_0^1 \mathcal{I}(\bar{k}; \gamma(R_{12}), \tau(R_{12}), \eta) \, \mathrm{d}\bar{k} \right|}, 
\qquad \text{with} \qquad
\delta \mathcal{I}(\bar{k}; \gamma(R_{12}), \tau(R_{12}), \eta) = \mathcal{I}(\bar{k}; \gamma, \tau, \eta) - \mathcal{I}^{(l_1,l_2)}_{\text{appr}}(\bar{k}; \gamma, \tau, \eta).
\end{equation}
The maximum value of this observable over the interval \([0,1]\) for \(\bar{k}\) provides an upper bound on the relative error introduced by truncating the Taylor expansion at a given order in \(\gamma\) and \(\tau\), in accordance with the mean value theorem for integrals. 
For instance,
truncating the expansion at the order $m_{\gamma}=3$ and $\tau=7$ we obtain
\begin{equation}
\begin{aligned}
& \mathcal{I}(\bar{k}; \gamma(R_{12}), \tau(R_{12}), \eta) \simeq \gamma \bigl(
- \dfrac{x^{4}\tau^{2}}{15\,(k_{Q}+x)^{2}}
+ \dfrac{x^{6}\tau^{4}}{210\,(k_{Q}+x)^{2}}
- \dfrac{x^{8}\tau^{6}}{7560\,(k_{Q}+x)^{2}}
+ \mathcal{O}(\tau^{7})
\bigr)
\\[6pt]
&\quad
+ \gamma^{3} \bigl(
- \dfrac{x^{2}(5k_{Q}^{2}+4xk_{Q}+x^{2})}{4\,(k_{Q}+x)^{4}}
+ \dfrac{x^{4}(5k_{Q}^{2}+4xk_{Q}+x^{2})\tau^{2}}{60\,(k_{Q}+x)^{4}}
- \dfrac{x^{6}(5k_{Q}^{2}+4xk_{Q}+x^{2})\tau^{4}}{3360\,(k_{Q}+x)^{4}}
+ \mathcal{O}(\tau^{7})
\bigr)
\\[6pt]
&\quad
+ \gamma^{5} \bigl(
- \dfrac{5x^{2}(63k_{Q}^{4}+122xk_{Q}^{3}+102x^{2}k_{Q}^{2}+42x^{3}k_{Q}+7x^{4})}{64\,(k_{Q}+x)^{6}}
+ \dfrac{7x^{4}(63k_{Q}^{4}+122xk_{Q}^{3}+102x^{2}k_{Q}^{2}+42x^{3}k_{Q}+7x^{4})\tau^{2}}{960\,(k_{Q}+x)^{6}}
\\[4pt]
&\qquad\qquad
- \dfrac{13x^{6}(63k_{Q}^{4}+122xk_{Q}^{3}+102x^{2}k_{Q}^{2}+42x^{3}k_{Q}+7x^{4})\tau^{4}}{53760\,(k_{Q}+x)^{6}}
+\\
&+\dfrac{x^{8}(63k_{Q}^{4}+122xk_{Q}^{3}+102x^{2}k_{Q}^{2}+42x^{3}k_{Q}+7x^{4})\tau^{6}}{241920\,(k_{Q}+x)^{6}}
+ \mathcal{O}(\tau^{7})
\bigr)
\\[6pt]
\end{aligned}
\label{eq:series3}
\end{equation}
We remark that 
\begin{itemize}
    \item The dependency on the mutual distances $R_{12}$ is fully
    encoded in $\gamma\propto R_{12}^{-3}$ and $\tau\propto R_{12}$;
    \item For any given order $l_{\gamma}$ there exists $l_{\tau}> 3 l_{\gamma}$ such that positive powers of $R_{12}$ can be generated;
    \item The term leading to $R^{-1}_{12}$ dependency comes uniquely 
    from $l_{\gamma}=1$ and $l_{\tau}=2$. This means that such an 
    interaction \textit{requires to take into account for 
    retardation effects}.
\end{itemize}

Based on this analysis, we find that choosing \(m_\gamma = 3\) and \(m_\tau = 3\) yields a reasonable approximation for the values of \(k_M\) considered in the main text (\(\eta \leq 2 \times 10^{-2}\)). Making explicit the dependency on $R_{12}$ of $\tau$ and $\gamma$ and recasting the terms of the expansion in powers of $R_{12}$ we obtain an approximated expression 
for the interaction potential
\begin{equation}
\label{eq:approx_integr_R}
     \mathcal{I}^{(3,3)}_{\text{appr}}(\bar{k},\gamma(R_{12}),\tau(R_{12}),\eta)= \left(\dfrac{c^{\rm int}_{1}(\bar{k})}{R_{12}}+\dfrac{c^{\rm int}_{7}(\bar{k})}{R_{12}^7}+\dfrac{c^{\rm int}_{9}(\bar{k})}{R_{12}^9}\right)
\end{equation}
where the coefficient reads as 
\begin{align}
    & \bar{c}^{(\mathcal{I})}_{1}(\bar{k}) = -\frac{\mathcal{A}_{0} k_M^2 \bar{k}^4}{15 (\bar{k}+\bar{k}_{Q})^2}\\
    & \bar{c}^{(\mathcal{I})}_{7}(\bar{k}) = \frac{\mathcal{A}_{0}^3 \bar{k}^4 k_M^2 \left(\bar{k}^2+4 \bar{k} \bar{k}_{Q}+5 \bar{k}_{Q}^2\right)}{60 (\bar{k}+\bar{k}_{Q})^4}\\
    & \bar{c}^{(\mathcal{I})}_{9}(\bar{k}) = -\frac{\mathcal{A}_{0}^3 \bar{k}^2 \left(\bar{k}^2+4 \bar{k} \bar{k}_{Q}+5 \bar{k}_{Q}^2\right)}{4 (\bar{k}+\bar{k}_{Q})^4}
\end{align}
After integrating the expression in Eq.\eqref{eq:approx_integr_R} with respect to $\bar{k}$, we obtain 
\begin{equation}
    \Delta V^{\rm int}_{12} = \Delta V\left[ \dfrac{\bar{c}_{1}}{R_{12}}+\dfrac{\bar{c}_{7}}{R_{12}^7}+\dfrac{\bar{c}_{9}}{R_{12}^9} \right]\,\,.
\end{equation}
where the coefficient reads
\begin{align}
\label{eq:coefficients_final}
    & \bar{c}_{1} =
\frac{\mathcal{A}_0 \, k_M^2 \left[ 2 (6 \bar{k}_Q^2 + 3 \bar{k}_Q - 1) \bar{k}_Q + 1 -24 (\bar{k}_Q + 1) \bar{k}_Q^3 \, \coth^{-1}(2 \bar{k}_Q + 1)\right]}{45 (\bar{k}_Q + 1)}
 \\
& \bar{c}_{7} = -\frac{\mathcal{A}_0^3 \, k_M^2 \, (3 \bar{k}_Q + 1)}{180 (\bar{k}_Q + 1)^3} \\
& \bar{c}_{9} = \frac{\mathcal{A}_0^3 (3 + 5 \bar{k}_Q)}{12 (1 + \bar{k}_Q)^3} \\
\end{align}
For all the choices of $k_M$ and the QDOs parameters that reproduce the Ar atom electric response properties, $c_{1}>0$ so that the $1/R$ term is repulsive.\\
We notice here that the coefficient in Eq.\eqref{eq:coefficients_final} are expressed as a function
of the QDO polarizability $\mathcal{A}_{0}$.
However, the polarizability can be expressed as a function 
of the fine structure constant $\alpha_{\rm fsc}$ as $\mathcal{A}_{0}=\alpha_{\rm fsc} \hbar c/m\omega^2$ so 
that the coefficients can be rewritten as follows 
in different system of units:
\begin{itemize}
    \item In SI units
    \begin{align}
\label{eq:coefficients_final_SIu}
    & \bar{c}_{1} = \alpha_{\rm fsc} \dfrac{\hbar c}{m\omega^2}
\frac{ k_M^2 \left[ 2 (6 \bar{k}_Q^2 + 3 \bar{k}_Q - 1) \bar{k}_Q + 1 -24 (\bar{k}_Q + 1) \bar{k}_Q^3 \, \coth^{-1}(2 \bar{k}_Q + 1)\right]}{45 (\bar{k}_Q + 1)}
 \\
& \bar{c}_{7} = -\alpha_{\rm fsc}^3 \left(\dfrac{\hbar c}{m\omega^2}\right)^3\frac{ \, k_M^2 \, (3 \bar{k}_Q + 1)}{180 (\bar{k}_Q + 1)^3} \\
& \bar{c}_{9} = \alpha_{\rm fsc}^3 \left(\dfrac{\hbar c}{m\omega^2}\right)^3 \frac{\, k_M^2 \, (3 \bar{k}_Q + 1)}{180 (\bar{k}_Q + 1)^3} \\
\end{align}
\item In natural units
\begin{align}
\label{eq:coefficients_final_nu}
    & \bar{c}_{1} = \alpha_{\rm fsc} \dfrac{1}{m\omega^2}
\frac{ k_M^2 \left[ 2 (6 \bar{k}_Q^2 + 3 \bar{k}_Q - 1) \bar{k}_Q + 1 -24 (\bar{k}_Q + 1) \bar{k}_Q^3 \, \coth^{-1}(2 \bar{k}_Q + 1)\right]}{45 (\bar{k}_Q + 1)}
 \\
& \bar{c}_{7} = -\alpha_{\rm fsc}^3 \left(\dfrac{1}{m\omega^2}\right)^3\frac{ \, k_M^2 \, (3 \bar{k}_Q + 1)}{180 (\bar{k}_Q + 1)^3} \\
& \bar{c}_{9} = \alpha_{\rm fsc}^3 \left(\dfrac{1}{m\omega^2}\right)^3 \frac{\, k_M^2 \, (3 \bar{k}_Q + 1)}{180 (\bar{k}_Q + 1)^3} \\
\end{align}
\item in atomic units
\begin{align}
\label{eq:coefficients_final_au}
    & \bar{c}_{1} = \dfrac{1}{m\omega^2}
\frac{ k_M^2 \left[ 2 (6 \bar{k}_Q^2 + 3 \bar{k}_Q - 1) \bar{k}_Q + 1 -24 (\bar{k}_Q + 1) \bar{k}_Q^3 \, \coth^{-1}(2 \bar{k}_Q + 1)\right]}{45 (\bar{k}_Q + 1)}
 \\
& \bar{c}_{7} = - \left(\dfrac{1}{m\omega^2}\right)^3\frac{ \, k_M^2 \, (3 \bar{k}_Q + 1)}{180 (\bar{k}_Q + 1)^3} \\
& \bar{c}_{9} =  \left(\dfrac{1}{m\omega^2}\right)^3 \frac{\, k_M^2 \, (3 \bar{k}_Q + 1)}{180 (\bar{k}_Q + 1)^3} \\
\end{align}
\end{itemize}

\begin{figure}
    \centering
    \includegraphics[width=0.45\linewidth]{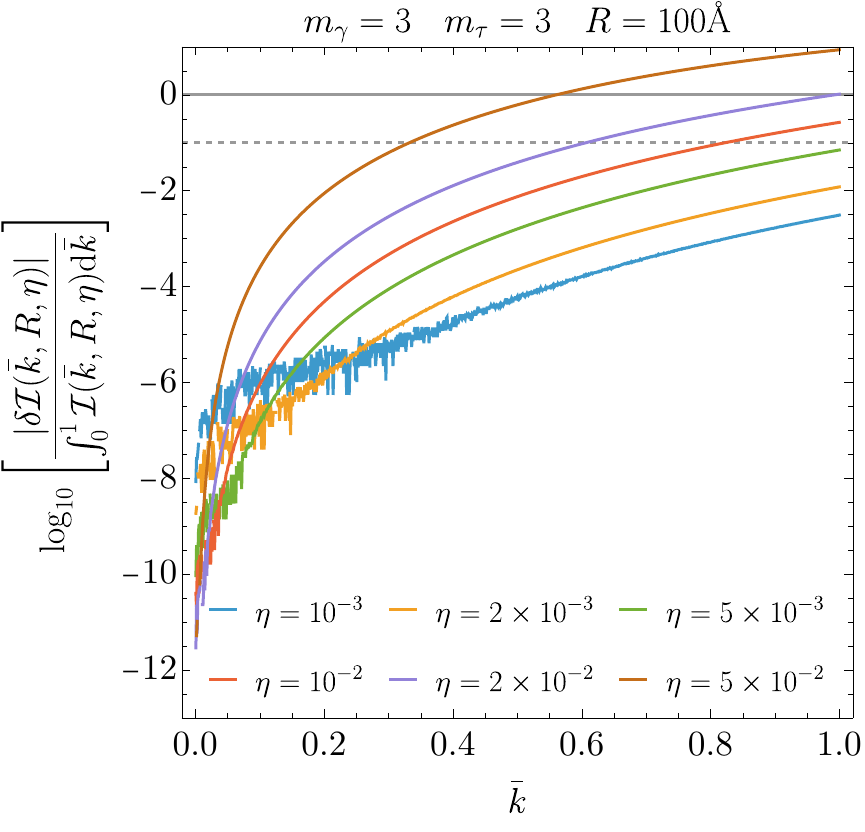}
    \hspace{0.5cm}
    \includegraphics[width=0.45\linewidth]{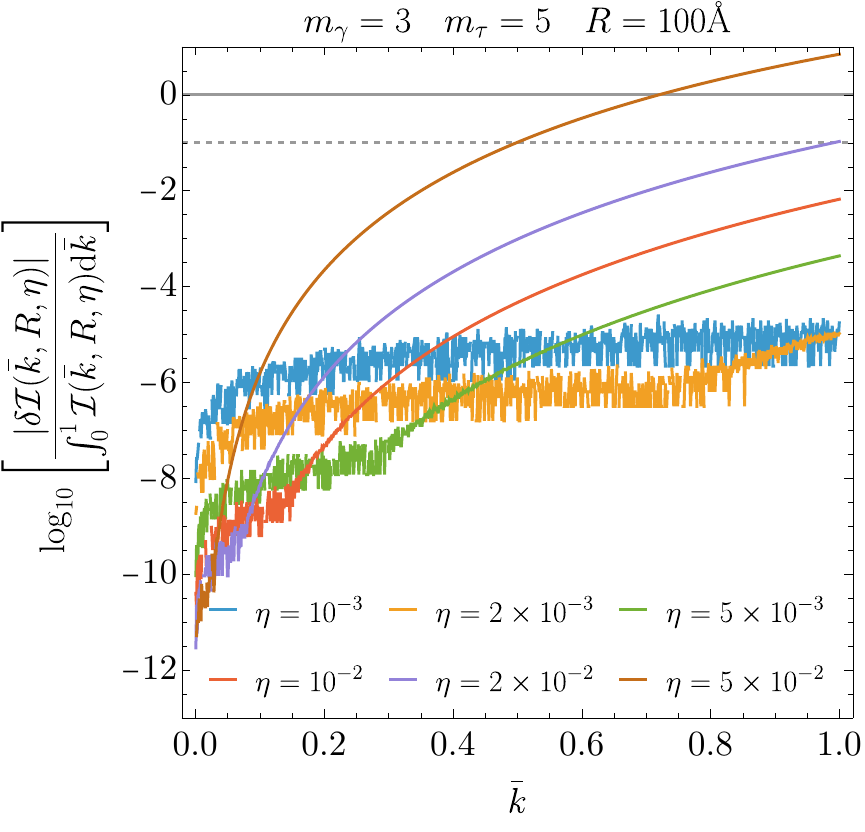}
    \caption{\textbf{Error analysis in the approximation of $\mathcal{I}_{\rm appr}(\bar{k};\gamma,\tau,\eta)$}. }
    \label{fig:enter-label}
\end{figure}

\section{Energy-Shift due to interaction Hamiltonian involving $A^2$}\label{sec:energy-shift-A2}
The interaction term $H_{\rm int}^{(2)}=\sum_{a=1,2}\frac{q^2}{2m} A^2(\bm{R}_a)$ represents two-photon processes, in which the interaction occurs through the simultaneous exchange of two photons between two centers. Since the vector potential $A$ is linear in terms of creation and annihilation operators of photons ($a^\dagger_{k,\lambda}, a_{k,\lambda}$), $A^2$ contains four types of combinations of $a_{k,\lambda}$ and $a^\dagger_{k,\lambda}$, namely 
$a^\dagger_{k,\lambda} a^\dagger_{k,\lambda}$,~ 
$a^\dagger_{k,\lambda} a_{k,\lambda}$,~ 
$a_{k,\lambda} a^\dagger_{k,\lambda}$, and 
$a_{k,\lambda} a_{k,\lambda}$. 
Therefore, the first order of perturbation merely contributes to self-energies that are independent of $R$. 
From the second-order perturbation, we obtain the leading-order energy shift due to the matter-field interaction Hamiltonian $H_{\rm int}^{(2)}$ as follows:
\begin{gather}
\label{2ndpert-A2-1}
\Delta E_2 = \sum_{I\neq 0}\frac{\langle 0|H_{int}^{(2)}|I\rangle \langle I|H_{int}^{(2)}|0\rangle}{E_0-E_I}\ ,
\end{gather}
where in the intermediate state $|I\rangle$, the field has to be in a two-photon excited state $|1_{k,\lambda}\, 1_{k',\lambda'}\rangle$, and the molecular state (MBD state) has to be in the ground state in order to result in non-vanishing energy shifts, hence, $E_0-E_I=-\hbar c(k+k')$. Therefore we have 
\begin{align}
\langle I|H_{int}^{(2)}|0\rangle 
&= \langle I|~ \sum_{a=1,2}\sum_{k,\lambda}\sum_{k',\lambda'} 
\frac{q_a^2}{2m_a}\frac{\hbar}{2 \epsilon_0 c V}
\frac{1}{\sqrt{kk'}}[\bm{e}(k,\lambda)\cdot\bm{e}(k',\lambda')]a^\dagger_{k,\lambda} a^\dagger_{k',\lambda'} e^{-i(\bm{k}+\bm{k}')\cdot\bm{R}_a} ~|0\rangle 
\nonumber\\
&=\sum_{a=1,2}\sum_{k,\lambda}\sum_{k',\lambda'} \frac{q_a^2}{2m_a} \frac{\hbar}{2 \epsilon_0 c V}
\frac{1}{\sqrt{kk'}}[\sum_{i=x,y,z} e_i(k,\lambda)e_i(k',\lambda')] e^{-i(\bm{k}+\bm{k}')\cdot\bm{R}_a}\ ,
\end{align}
thus
\begin{align}
\label{2ndpert-A2-2}
\Delta E_2 &= \sum_{I\neq 0}\frac{\langle 0|H_{int}^{(2)}|I\rangle \langle I|H_{int}^{(2)}|0\rangle}{E_0-E_I}
=-\sum_{a,b=1}^{2}\sum_{i,j=x,y,z}\sum_{k,k'} 
\frac{q_a^2 \, q_b^2}{4 m_a m_b} 
\left(\frac{\hbar}{2 \epsilon_0 c V}\right)^2
\frac{1}{\hbar c\, kk'(k+k')} e^{i(\bm{k}+\bm{k}')\cdot(\bm{R}_b-\bm{R}_a)}
\nonumber\\
&\hspace{8cm}\times
[\sum_{\lambda=1}^{2} e_i(k,\lambda)e_j(k,\lambda)]\, [\sum_{\lambda'=1}^{2}e_i(k',\lambda')e_j(k',\lambda')]
\nonumber\\
&=-\sum_{a,b=1}^{2}\sum_{i,j=x,y,z}\sum_{k,k'} 
\frac{q_a^2 \, q_b^2}{4 m_a m_b} 
\left(\frac{\hbar}{2 \epsilon_0 c V}\right)^2
\frac{1}{\hbar c\, kk'(k+k')} e^{i(\bm{k}+\bm{k}')\cdot(\bm{R}_b-\bm{R}_a)}
[\delta_{ij}-\hat{k}_i\hat{k}_j]\, [\delta_{ij}-\hat{k}'_i\hat{k}'_j]\ ,
\end{align}
where in the last line of \eqref{2ndpert-A2-2} the expressions inside the square brackets result from summing over polarizations of the field. 
In the continuum limit of the quantization volume $V$, the sums over the wavenumbers $k$ and $k'$ are replaced by integrals, {\it i.e} $\frac{1}{V}\sum_k\rightarrow \int \frac{d^3k}{(2\pi)^3}$. Using the spherical coordinates for the integrals over the wavevectors $\bm{k}$ and $\bm{k}'$, we arrive at 
\begin{align}
\label{2ndpert-A2-3}
\Delta E_2 &=
-\sum_{a,b=1}^{2}\sum_{i,j=x,y,z} 
\frac{q_a^2 \, q_b^2}{4 m_a m_b} 
\frac{\hbar}{4 \epsilon_0^2 c^3}
\left\{\frac{1}{8\pi^3}\int_0^\infty k\,dk \int e^{i\bm{k}\cdot(\bm{R}_b-\bm{R}_a)} [\delta_{ij}-\hat{k}_i\hat{k}_j] d\Omega\right.
\nonumber\\
&\hspace{5.3cm}
\left.\left\{\frac{1}{8\pi^3}\int_0^\infty \frac{k'}{(k+k')}dk' \int e^{i\bm{k}'\cdot(\bm{R}_b-\bm{R}_a)}[\delta_{ij}-\hat{k}'_i\hat{k}'_j] d\Omega'\right\}\right\}
\ .
\end{align}
When $a=b$, the virtual transitions of the total system in the two-step process of the second-order perturbation correspond to the creation and annihilation of a pair of virtual photons at the same center, hence giving an energy shift that is independent of the molecular state and can be considered as a constant shift to all energy levels. When $a\neq b$, there are two possibilities: $(a,b)=(1,2)$ that indicates creation of two virtual photons at $\bm{R}_1$ and annihilation of them at $\bm{R}_2$, or $(a,b)=(2,1)$ for the opposite creation/annihilation process. The energy shifts resulting from such processes depend on the distance between the photon-creation/annihilation centers, hence depend on the molecular state. Therefore, for $a\neq b$ we have $\bm{R}_b-\bm{R}_a=\pm\bm{R}$ where $\bm{R}=\bm{R}_2-\bm{R}_1$, and we end up with the known angular integral 
\begin{align}
\tau_{ij}(kR) = \frac{1}{4\pi}\int e^{\pm i\bm{k}\cdot\bm{R}}[\delta_{ij}-\hat{k}_i\hat{k}_j] d\Omega = 
(\delta_{ij}-R_i\,R_j)\frac{\sin(kR)}{kR} + 
(\delta_{ij}-3R_i\,R_j)\left[\frac{\cos(kR)}{k^2R^2}-\frac{\sin(kR)}{k^3R^3}\right]\ .
\end{align}
Using this integral identity expression~\eqref{2ndpert-A2-3} reduces to 
\begin{align}
\label{2ndpert-A2-4}
\Delta E_2 &=
-2\sum_{i,j=x,y,z} 
\frac{q_1^2 \, q_2^2}{4 m_1 m_2} 
\frac{\hbar}{16\pi^4 \epsilon_0^2 c^3}
\left\{\int_0^\infty k\, \tau_{ij}(kR) \,dk \right.
\left. \left\{\int_0^\infty \frac{k'}{(k+k')} \tau_{ij}(k'R)\, dk'\right\}\right\}
\ .
\end{align}
To decouple the integrals, we use the integral identity 
\begin{equation}
\frac{1}{k+k'}=\int_0^\infty e^{-u(k+k')}\,du \ ,
\end{equation}
hence equation \eqref{2ndpert-A2-4} becomes
\begin{align}
\label{2ndpert-A2-5}
\Delta E_2 &=
-2\sum_{i,j=x,y,z} 
\frac{q_a^2 \, q_b^2}{4 m_a m_b} 
\frac{\hbar}{16\pi^4 \epsilon_0^2 c^3}
\int_0^\infty \left\{
\left[\int_0^\infty k\, \tau_{ij}(kR)\, e^{-uk} \,dk \right]
\left[\int_0^\infty k'\,\tau_{ij}(k'R)\, e^{-uk'}\, dk'\right]\right\} du
\nonumber\\
&=
-\frac{q_a^2 \, q_b^2}{2 m_a m_b} 
\frac{\hbar}{16\pi^4 \epsilon_0^2 c^3}
\sum_{i,j=x,y,z} 
\int_0^\infty 
\left[ F_{ij}(u,R) \right]^2\, du
\ , 
\end{align}
where
\begin{align}
\label{2ndpert-A2-k-integral-1}
F_{ij}(u,R)&=\int_0^\infty k\, \tau_{ij}(kR)\, e^{-uk} \,dk
\nonumber\\
&=
\int_0^\infty e^{-uk}\, 
\left\{(\delta_{ij}-R_i\,R_j)\frac{\sin(kR)}{R} + 
(\delta_{ij}-3R_i\,R_j)\left[\frac{\cos(kR)}{k R^2}-\frac{\sin(kR)}{k^2R^3}\right]\right\}
\,dk
\nonumber\\
&=
\frac{(\delta_{ij}-R_i\,R_j)}{u^2+R^2} +
\frac{(\delta_{ij}-3R_i\,R_j)}{R}
\int_0^\infty e^{-uk}
\left[\frac{\cos(kR)}{k R}-\frac{\sin(kR)}{k^2R^2}\right]
\,dk \ .
\end{align}
Taking the remaining integral in the definition of $F_{ij}(u,R)$ is challenging. We notice that the integrand in the last line of equation~\eqref{2ndpert-A2-k-integral-1} is, apart from the exponential part, a first-order derivative with respect to $k$, {\it i.e.}
\begin{equation}
\left[\frac{\cos(kR)}{kR}-\frac{\sin(kR)}{k^2 R^2}\right]=\frac{1}{R}\frac{d}{dk}\left(\frac{\sin(kR)}{k R}\right) \ .
\end{equation}
Using the above equality, we calculate the integral as follows
\begin{align}
\label{2ndpert-A2-k-integral-2}
&\frac{(\delta_{ij}-3R_i\,R_j)}{R}
\int_0^\infty e^{-uk}
\left[\frac{\cos(kR)}{k R}-\frac{\sin(kR)}{k^2R^2}\right]
\,dk 
=
\frac{(\delta_{ij}-3R_i\,R_j)}{R^2}
\int_0^\infty e^{-uk}
\left[\frac{d}{dk}\left(\frac{\sin(kR)}{k R}\right)\right]
\,dk 
\nonumber\\
=&
\frac{(\delta_{ij}-3R_i\,R_j)}{R^2}
\left[e^{-uk} \frac{\sin(kR)}{kR}\right]_0^\infty +\frac{(\delta_{ij}-3R_i\,R_j)}{R^2}\int_0^\infty \frac{\sin(kR)}{kR}ue^{-uk}\,dk
\nonumber\\
=&
(\delta_{ij}-3R_i\,R_j)\left[-\frac{1}{R^2} + \frac{u}{R^3}\arctan(\frac{R}{u})\right]\ .
\end{align}
Therefore, $F_{ij}(u,R)$ is obtained as
\begin{align}
\label{2ndpert-A2-k-integral-3}
F_{ij}(u,R)&=
\frac{(\delta_{ij}-R_i\,R_j)}{u^2+R^2} +
(\delta_{ij}-3R_i\,R_j)\left[-\frac{1}{R^2} + \frac{u}{R^3}\arctan(\frac{R}{u})\right]\ .
\end{align}

As before, we assume the two QDOs are placed on the $z$-axis such that $\bm{R}_2-\bm{R}_1=\bm{R}=R\hat{\bm{z}}$. With this assumption, we perform the sums over $i$ and $j$ to obtain 
\begin{align}
\label{2ndpert-A2-k-integral-4}
\sum_{i,j=x,y,z} \left[F_{ij}(u,R)\right]^2 
&=
\frac{6}{R^4}
-\frac{4}{R^2 \left(R^2+u^2\right)}
+\frac{2}{(R^2+u^2)^2}
-\frac{12 u }{R^5}\arctan\left(\frac{R}{u}\right)
+\frac{4 u }{R^3 \left(R^2+u^2\right)}\arctan\left(\frac{R}{u}\right)
\nonumber\\
&
+\frac{6 u^2 }{R^6}\arctan\left(\frac{R}{u}\right)^2 \ .
\end{align}
Replacing the sum~\eqref{2ndpert-A2-k-integral-4} into~\eqref{2ndpert-A2-5} and taking the $u$-integral we arrive at the interaction energy
\begin{align}
\label{2ndpert-A2-6}
\Delta E_2 &=
-\frac{q_a^2 \, q_b^2}{2 m_a m_b} 
\frac{\hbar}{16\pi^4 \epsilon_0^2 c^3}
\frac{\pi}{2R^3} \ .
\end{align}

\end{document}